\documentclass[prb,twocolumn,showpacs,amsmath]{revtex4}
\usepackage{graphicx}
\usepackage{dcolumn}
\usepackage{bm}
\usepackage{booktabs}

\begin{document}

\preprint{FLEX}

\title{Phase diagram and Gap anisotropy in Iron-Pnictide Superconductors}

\author{Hiroaki Ikeda$^1$}%
\email{hiroaki@scphys.kyoto-u.ac.jp}
\author{Ryotaro Arita$^{2,3,4}$}%
\author{Jan Kune\v{s}$^{5,6}$}%
\affiliation{$^1$Department of Physics, Kyoto University, Kyoto, 606-8502, Japan\\
$^2$Department of Applied Physics, University of Tokyo, Tokyo 113-8656, Japan\\
$^3$TRIP, JST, Sanbancho, Chiyoda, Tokyo 102-0075, Japan\\
$^4$CREST, JST, Hongo, Tokyo 113-8656, Japan\\
$^5$Center for Electronic Correlations and Magnetism, Theoretical Physics III, University of Augsburg, 86135 Augsburg, Germany \\
$^6$Institute of Physics,
Academy of Sciences of the Czech Republic, Cukrovarnick\'a 10,
162 53 Praha 6, Czech Republic
}%

\date{\today}%

\begin{abstract}
Using the fluctuation-exchange approximation, we study an effective five-band Hubbard model for iron-pnictide superconductors obtained from the first-principles band structure. 
We preclude deformations of the Fermi surface due to electronic correlations by introducing a static potential, which mimics the effect of charge relaxation. 
Evaluating the Eliashberg equation for various dopings and interaction parameters, we find that superconductivity can sustain higher hole than electron doping.
Analyzing the symmetry of the superconducting order parameter we observe clear differences between the hole and electron-doped systems. 
We discuss the importance of the pnictogen height for superconductivity.
Finally, we dissect the pairing interaction into various contributions, which allows us to clarify the relationship between the superconducting transition temperature and the proximity to the anti-ferromagnetic phase. 
\end{abstract}

\pacs{
}

\maketitle

\section{Introduction}
The recent discovery of superconductivity at $26$K in LaFeAsO$_{1-x}$F$_x$ opened a new field of a highly intensive research in the material science.~\cite{rf:Kamihara}
In a short period of time, the superconducting transition temperature $T_c$ has been elevated to over 50~K by substitution of another rare earths for La, 
which yields the highest $T_c$ outside cuprates.~\cite{rf:Chen,rf:Ren,rf:Cheng}
At present, there exist the 1111 systems represented by LaFeAsO, the 122 systems with BaFe$_2$As$_2$,~\cite{rf:Rotter} the 111 systems with LiFeAs,~\cite{rf:Wang} and the 11 systems with Fe(Se,Te).~\cite{rf:Hsu,rf:Yeh}
These four families hold similar Fe-pnictogen layers, and are supposed to possess the same superconducting pairing mechanism.~\cite{rf:Ishida}
The superconducting phase appears in a close proximity to the stripe-type anti-ferromagnetic (AF) phase of the undoped systems.
Early on, it was argued that the AF spin fluctuations originate from the nesting between the two-dimensional cylindrical Fermi surfaces (two hole surfaces around $\varGamma$ point and two electron surfaces around $M$ point), and that they give rise to the sign-reversing $s$-wave ($s_\pm$) superconducting state.~\cite{rf:Mazin}
Kuroki {\it et al.} constructed an effective five-band model Hamiltonian in the unfolded Brillouin zone (BZ), which can describe the band structure of LaFeAsO near the Fermi level, and analyzed it within the random-phase approximation (RPA) obtaining a similar pairing state.~\cite{rf:Kuroki}
This conclusion was confirmed within the RPA studies,~\cite{rf:Yanagi,rf:Graser} the third-order perturbation theory,~\cite{rf:Nomura} the functional renormalization group,~\cite{rf:DHLee} and the fluctuation-exchange (FLEX) approximation.~\cite{rf:Yao,rf:Ikeda}
The $s_\pm$-wave state has been actively discussed as a promising candidate for the pairing symmetry in the iron-pnictide superconductors.

The undoped LaFeAsO parent compound has a stripe-type AF ground state. 
With the electron doping due to substitution of F for O, the AF phase abruptly vanishes in a first order way, and the superconducting phase appears.
The transition temperature barely changes with further the electron doping over the range $x=0.04 - 0.14$ for LaFeAsO$_{1-x}$F$_x$.
The NMR-$1/T_1$ study in the normal state revealed that the strong AF fluctuations in the undoped case are dramatically suppressed with electron doping, 
and a pseudogap behavior was observed above $x=0.1$.~\cite{rf:Nakai,rf:Grafe,rf:Ahilan,rf:Mukuda}
Such a pseudogap behavior has been observed also with the photoemission spectroscopy.~\cite{rf:TSato}
This implies that the pairing mechanism cannot be attributed solely to the AF spin fluctuations.
The NMR-$1/T_1$ relaxation rate in the superconducting state follows the $T^3$ dependence,~\cite{rf:Nakai,rf:Grafe,rf:Mukuda} but recently, also a $T^6$-like behavior has been reported.~\cite{rf:Kobayashi}
Thus, it is not clear whether the power-law behavior reflects the superconducting node or not.
Rather, the lack of the residual density of states suggests a fully-gapped state with a gap minima.~\cite{rf:Nagai}

In BaFe$_2$As$_2$, the electron doping by substitution of Co for Fe and the hole doping by substitution of K for Ba are available.~\cite{rf:Rotter,rf:Sefat}
The superconductivity exists in a wider region for the hole doping than for the electron doping, and even the end material KFe$_2$As$_2$ is superconducting.~\cite{rf:Rotter2}
From the NMR-$1/T_1$ study in the normal state, the correlation between the AF spin fluctuations and $T_c$ can be deduced.~\cite{rf:Ning,rf:Fukazawa,rf:Matano,rf:Mukuda2}
As for the superconducting symmetry, in the hole-doped region, $T^5$-dependence of the NMR-$1/T_1$ (Ref.~\onlinecite{rf:Yashima}) and the exponential behavior of the penetration depth~\cite{rf:Hashimoto} indicate a fully-gapped superconductivity.
This is supported by the direct observation with the angle-resolved photoemission spectroscopy (ARPES).~\cite{rf:Ding,rf:Nakayama}
In the inelastic neutron scattering measurements, development of the resonance peak below $T_c$ was reported, although whether it means the sign change in the superconducting gap or not is not clear yet.~\cite{rf:Christianson,rf:Inosov,rf:Zhao,rf:Onari2}
In addition, quite recently, indications were found that KFe$_2$As$_2$ is a multi-gap system with line nodes,~\cite{rf:Fukazawa2,rf:Dong} and the As-P system shows a line-nodal behavior with high $T_c$.~\cite{rf:Hashimoto2,rf:Nakai2}

In FeSe, an enhancement of $T_c$ with pressure was reported, accompanied by an increase in the AF spin fluctuations.~\cite{rf:Imai,rf:Kotegawa}
The neutron scattering shows a correlation between the superconductivity and the stripe-type AF spin fluctuations, and below $T_c$, a development of a remarkable resonance peak.~\cite{rf:Qiu}
The thermal conductivity~\cite{rf:Dong2} and the scanning tunnel microscope~\cite{rf:Hanaguri} show a fully-gapped behavior.
Moreover, the phase-sensitive analysis seems to be consistent with the $s_\pm$-wave superconductivity.

Although the $s_\pm$-state is the prime candidate for explanation of pnictide superconductivity, whether the robustness of the superconductivity with respect to the presence of impurities can be understood 
within the $s_\pm$-scenario is a key issue for the future.~\cite{rf:Onari}
In addition, the recently discovered 42622 system with perovskite-block layer~\cite{rf:Ogino,rf:Zhu} seems to possess a considerably different band structure,~\cite{rf:Pickett,rf:Mazin3} while the $T_c$ is comparable with the other pnictides.
Whether this system possesses the same pairing mechanism is another open problem.

In order to understand the phase diagrams and the magnitude of $T_c$ of iron pnictides, it is necessary to study superconductivity and the correlation effects using realistic microscopic Hamiltonians.
The thermal-Hall conductivity~\cite{rf:Ong} and the microwave conductivity~\cite{rf:Hashimoto} in the superconducting state indicate a strong scattering between the quasiparticles, and 
the mass enhancement factor observed in the ARPES,~\cite{rf:Lu} the optical spectroscopy,~\cite{rf:Qazilbash} and the de Haas-van Alphen experiment~\cite{rf:Shishido} is as large as $2-3$.
In addition, {\it ab initio} estimates of the interaction parameters suggest that these systems are moderately, not weekly, correlated.~\cite{rf:Nakamura,rf:Miyake,rf:Anisimov}

In this paper, we investigate the superconductivity by including the correlation effect within the FLEX approximation.
As in the preceding study (Paper I),~\cite{rf:Ikeda} we encounter the following problem.
In the intermediate correlation regime, the renormalized band structure deviates drastically from the local-density approximation (LDA) one, which leads to substantial changes in the Fermi surface and the magnetic fluctuations, and thus spoils the good agreement of the LDA Fermi surface with ARPES data.
The effect was traced to a shift of the renormalized $d_{3z^2-r^2}$ site energy closer to the Fermi level, which leads to enlarging of the Fermi surface around the $\varGamma'$ in the unfolded BZ and shrinking of the other sheets. 
We believe, as discussed below, that it is an unphysical artifact of combining the {\it ab initio} band structure with the FLEX approximation.
In Paper I, as a tentative method, we shifted the site energy of $d_{3z^2-r^2}$ to preserve the shape of the Fermi surface in the renormalized band structure.
Although this allowed us to investigate the superconductivity with the effect of correlations, the value of the shift is ambiguous.
In Paper II,~\cite{rf:Arita} we constructed and studied an effective four-band model for $d_{xy}, d_{yz/zx}$ and $d_{x^2-y^2}$ assuming that $d_{3z^2-r^2}$ stays below the Fermi level
and is always irrelevant for the low-energy physics.
While the Fermi surface is not deformed so much even in the intermediate correlation regime as expected, the magnetic structure still drastically changes and the stripe-type AF never becomes dominant. 
This lead us to suggest that the high-energy physics such as the interactions between localized spins, which is not considered in the FLEX, 
may be important to understand the stripe-type AF in iron-pnictides.

In the present study, we examine the five-band model, and propose a simple way to avoid the drastic deformation of the Fermi surface by adding a static potential to the FLEX self-energy. 
We argue that such potential mimics the restoring force due to charge relaxation.
With this modification we obtain a strong stripe-type AF fluctuations also in the intermediate correlation regime.
Then, we evaluate the Eliashberg equation in the intermediate correlation regime, and investigate the phase diagram 
as a function of doping.
In addition, we clarify how to understand the correlation between the AF spin fluctuations and $T_c$, and where the paring glue for superconductivity comes from.

In the following section, first, we calculate the LDA band structure in LaFeAsO, and make {\it ab initio} construction of the effective five-band Hubbard model.
In Sec.\ref{sec:cFLEX} we demonstrate the results of the FLEX for the five-band Hubbard model, and discuss what kind of problems they have.
In Sec.\ref{sec:trouble} we introduce a simple way to avoid deformations of the Fermi surface due to electronic correlations, which mimics the effect of charge relaxation.
In Sec.\ref{sec:mFLEX} we verify that the modified FLEX well works even in the intermadiate correlation regime.
In Sec.\ref{sec:SC}, with this method we investigate the doping dependence of the eigenvalue in the Eliashberg equation.
We show that the obtained phase diagram qualitatively explains the overall feature in LaFeAsO and BaFe$_2$As$_2$, and then, the pnictogen height is important for high $T_c$.
In addition, we investigate the gap anisotropy of the pairing function obtained in several doping cases.
Finally, we suggest that it is oversimplified that the pairing mechanism in this system is attributed to only the conventional AF spin fluctuation.
In Appendices A$-$D, we summarize hopping integrals in the five-band model and the technical parts in symmetry consideration of Hamiltonian and FLEX calculations.

\section{Electronic Band Structure and Model Hamiltonian}
\begin{figure}
\centering
\vspace{5pt}
\includegraphics[width=65mm]{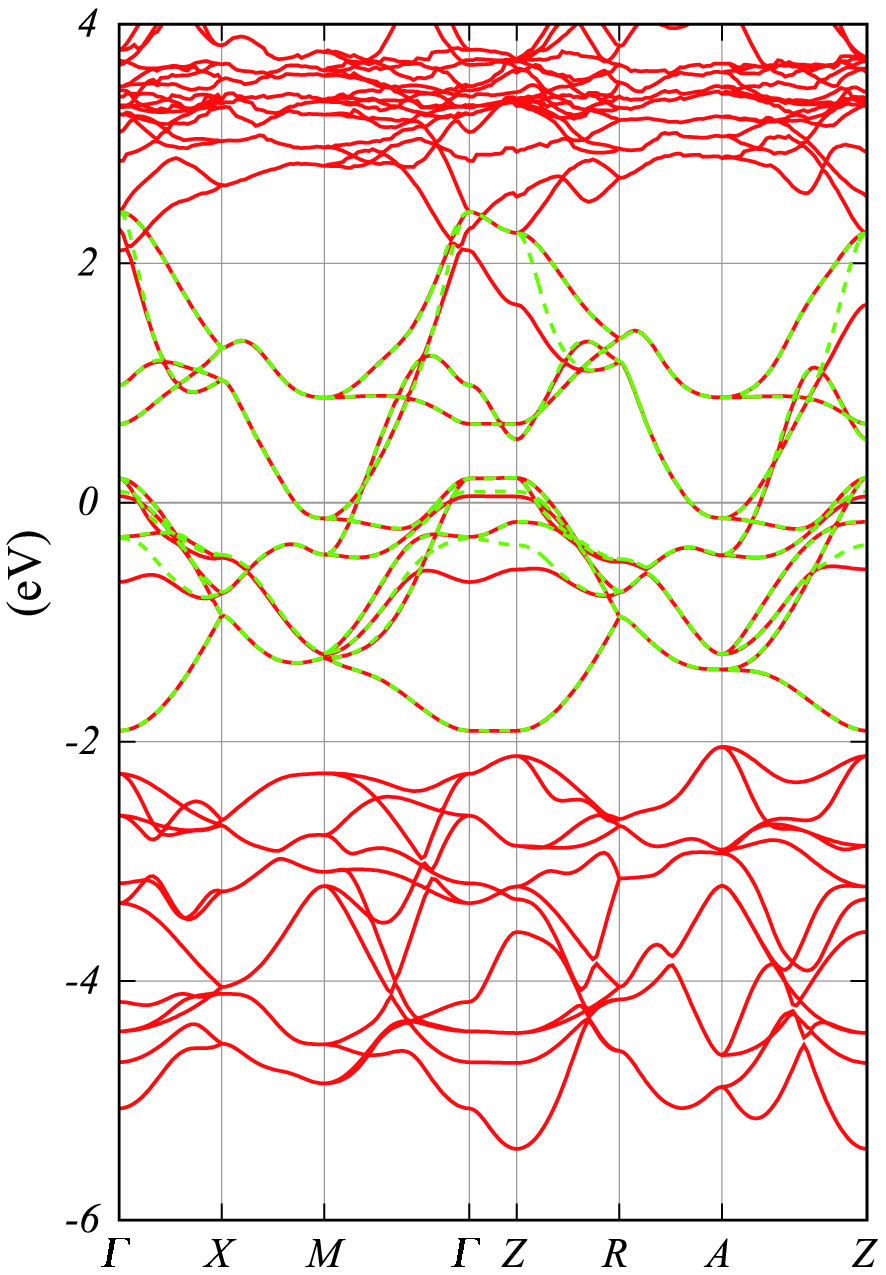}
\caption{(Color online) The electronic band structure of the undoped LaFeAsO. Red lines denote the band structure in the LDA calculation. Green dashed lines are the ten band model obtained with the MLWFs method.}
\label{fig:Disp}
\centering
\vspace{10pt}
\includegraphics[height=34mm]{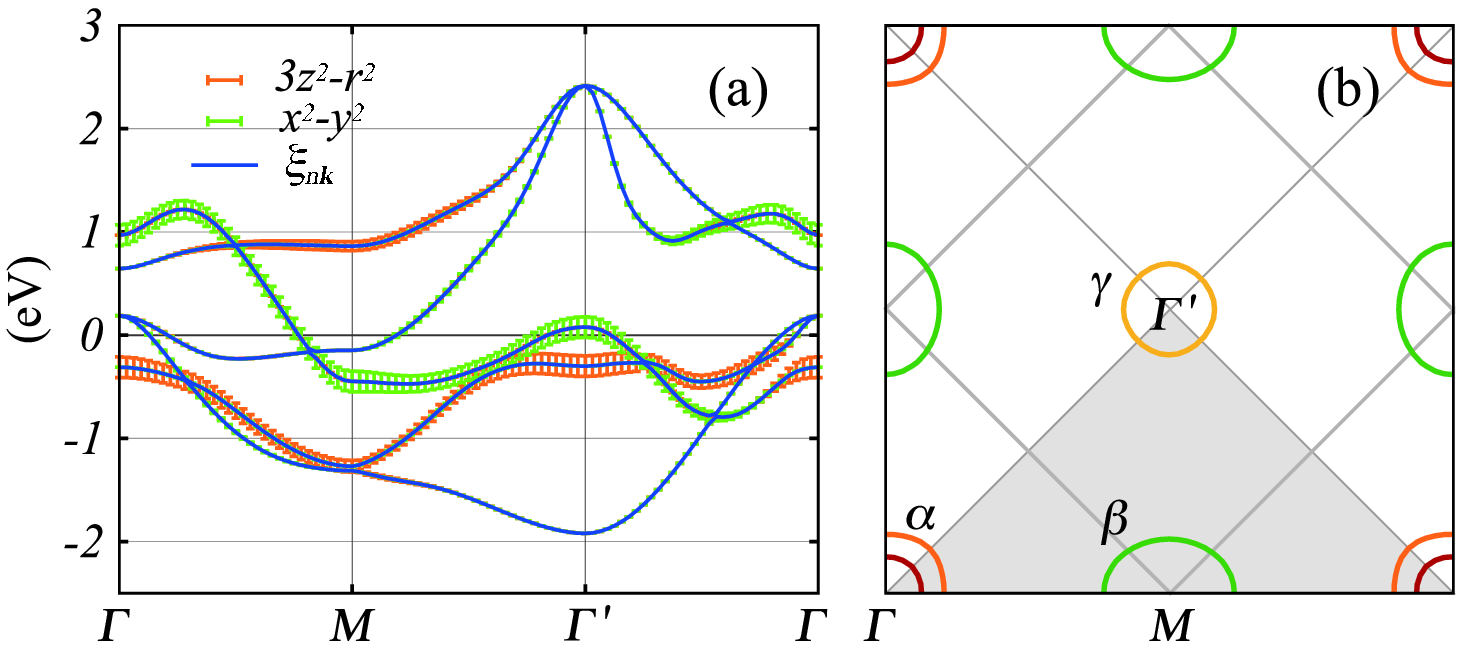}
\caption{(Color online) (a) The five-band structure and (b) The Fermi surface in the unfolded BZ. Orange and green bars in (a) represent weights of $3z^2-r^2$ and $x^2-y^2$ orbitals of Fe 3d, respectively. In (b), the inner square rotated 45 degrees denotes the folded original BZ. The shaded triangle stands for the irreducible part of the unfolded BZ.}
\label{fig:unfold}
\end{figure}
First, we carry out LDA band-structure calculation using WIEN2k package in the APW+local orbital basis.~\cite{rf:wien2k}
The band structure of iron-pnictides is known to be sensitive to the internal coordinate $z$, which determines distance of the pnictogen from the iron layer in the unit of the $c$ lattice parameter.
Experimentally, the superconducting transition temperature $T_c$ is well correlated with the Fe-$X$-Fe ($X$=pnictogen) bond-angle, which is related to the internal coordinate $z$.~\cite{rf:Lee}
Thus $z$ is an important parameter of the iron-pnictides.

In our calculations, performed for non-spin-polarized undoped LaFeAsO, we use
the experimental values for the crystallographic parameters [$a$=4.035~(\AA), $c$=8.741~(\AA), $z_{\rm La}=0.1415$, and $z_{\rm As}=0.6512$].~\cite{rf:Kamihara} 
In Fig.~\ref{fig:Disp}, we show the calculated band structure.
As $4p$ and O $2p$ hybrid bands are located between $-5\sim -2$eV, the Fe $3d$ orbitals are dominant contribution to the bands near the Fermi level, and the La $4f$ bands are situated around $3\sim 4$eV.
The Fermi surface consisting of quasi-two dimensional cylinders is essentially the same as the one obtained by Mazin {\it et al.}~\cite{rf:Mazin2}

Next, we use the recently developed interface~\cite{rf:Kunes} to wannier90 code~\cite{rf:MLWFs} and construct the maximally localized Wannier functions (MLWFs) spanning the Hilbert space of the Fe $3d$ bands.
The tight-binding model on the MLWFs basis provides an input for further calculations.
The tight-binding band structure (see Appendix \ref{sec:hopping}), marked by the green dashed line in Fig.~\ref{fig:Disp}, represents well the LDA Fe $3d$ bands.
Deviations around $\sim -0.5$ and $\sim 2$eV originate from strong hybridization with As $4p$, O $2p$ and La $5d$ orbitals.
The ten-band model can be unfolded~\cite{rf:Kuroki} to five bands in a doubled BZ, shown in 
Fig.~\ref{fig:unfold}. The band structure is similar to that of the preceding studies.~\cite{rf:Kuroki,rf:Ikeda}
The Fermi surface is composed of two hole sheets ($\alpha$) around the $\varGamma$ point, two electron sheets ($\beta$) around the $M$ point, and a hole sheet ($\gamma$) around the $\varGamma'$ point 
(the $\varGamma$ point of the original folded BZ).
The $\alpha$ surface and a part of the $\beta$ surface are dominated by the $d_{yz}$ and $d_{zx}$ character.
The $\gamma$ surface and another part of the $\beta$ surface arise from $d_{x^2-y^2}$ band.
The Fermi surface is characterized by nesting with $\bm{Q}=(\pi,0)$, which corresponds to stripe-type AF ordering.
In fact, the $(\pi,0)$ AF spin fluctuation is the dominant fluctuation within RPA.~\cite{rf:Kuroki}
This fluctuation is predominantly due to scattering between the $\beta$ and $\gamma$ surfaces with $d_{x^2-y^2}$ character.
Thus, the presence of the $\gamma$ surface with the high density of states is vital for this AF fluctuation.~\cite{rf:Ikeda,rf:Arita,rf:Kuroki2}
Moreover, since the $(\pi,0)$ AF fluctuations are considered the principal pairing glue for the sign-reversing $s_\pm$ pairing, 
the size of the $\gamma$ surface plays a key role for the high-$T_c$ superconductivity in this system.
However, it has been shown that the $\gamma$ hole surface is missing in the band structure in the iron-pnictides with perovskite-block layer~\cite{rf:Pickett,rf:Mazin3} mentioned in Sec. I.

Construction of the model Hamiltonian is completed by adding the on-site Coulomb interaction,
\begin{subequations}
\begin{alignat}{1}
H'&=\frac{U}{2}\sum_{i\ell}\sum_{\sigma}c^\dag_{i\ell\sigma}c^\dag_{i\ell\bar\sigma}c_{i\ell\bar\sigma}c_{i\ell\sigma} \\
&+\frac{U'}{2}\sum_{i\ell\ne m}\sum_{\sigma\sigma'}c^\dag_{i\ell\sigma}c^\dag_{im\sigma'}c_{im\sigma'}c_{i\ell\sigma} \\
&+\frac{J}{2}\sum_{i\ell\ne m}\sum_{\sigma\sigma'}c^\dag_{i\ell\sigma}c^\dag_{im\sigma'}c_{i\ell\sigma'}c_{im\sigma} \\
&+\frac{J'}{2}\sum_{i\ell\ne m}\sum_{\sigma}c^\dag_{i\ell\sigma}c^\dag_{i\ell\bar\sigma}c_{im\bar\sigma}c_{im\sigma},
\end{alignat}
\end{subequations}
where $\sigma=\pm$ and $\bar\sigma=-\sigma$, and $c^\dag_{i\ell\sigma}$, and $c_{i\ell\sigma}$ 
are the creation and annihilation operators in the  basis of real harmonics $3z^2-r^2$, $xz$, 
$yz$, $x^2-y^2$, and $xy$ located at Fe sites.
Since the Fermi surface of LaFeAsO is quasi-two-dimensional cylinder, for simplicity, we restrict ourselves to a two-dimensional $k_z=0$ space.
In addition, to meet the rotation invariance of the atomic orbitals in the orbital space, we take $U=U'+2J$ and $J=J'$.

In the following section, we show the results of the FLEX calculations.
The technical parts of FLEX are summarized in Appendix D.
In the actual calculations, we take $64\times 64$ meshes in the unfolded BZ and 1024 Matsubara frequencies.
In this case, we can safely carry out the FLEX calculation for $T\gtrsim 0.002$.
We set $T=0.003$ throughout this paper.
As the numerical analytic continuation, we use the Pad\'e approximation.

\section{Results for undoped case}
\subsection{Band-structure renormalization}\label{sec:cFLEX}
\begin{figure}
\centering
\vspace{5pt}
\includegraphics[width=85mm]{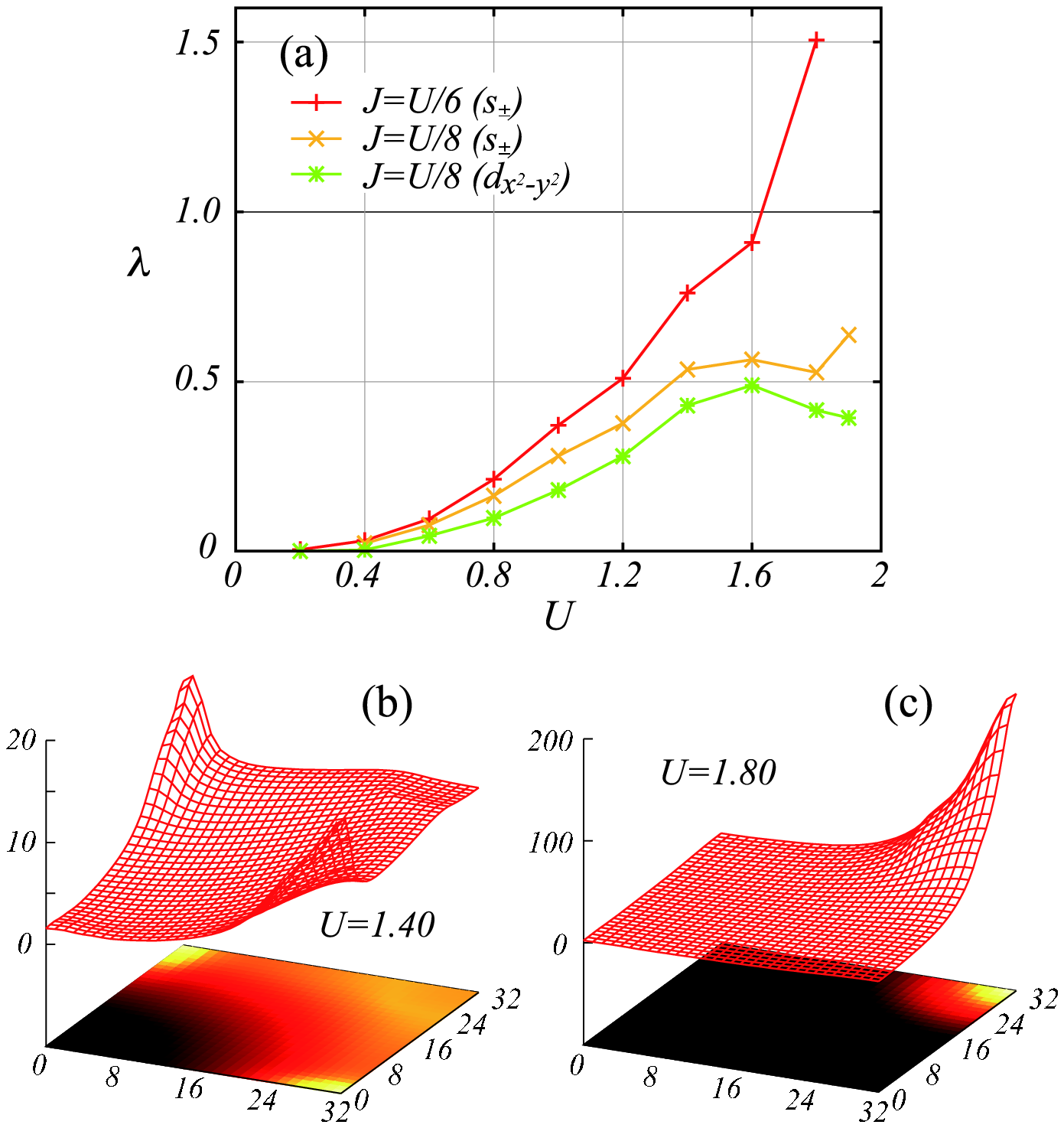}

\vspace{10pt}
\hspace{-4pt}\includegraphics[width=60mm]{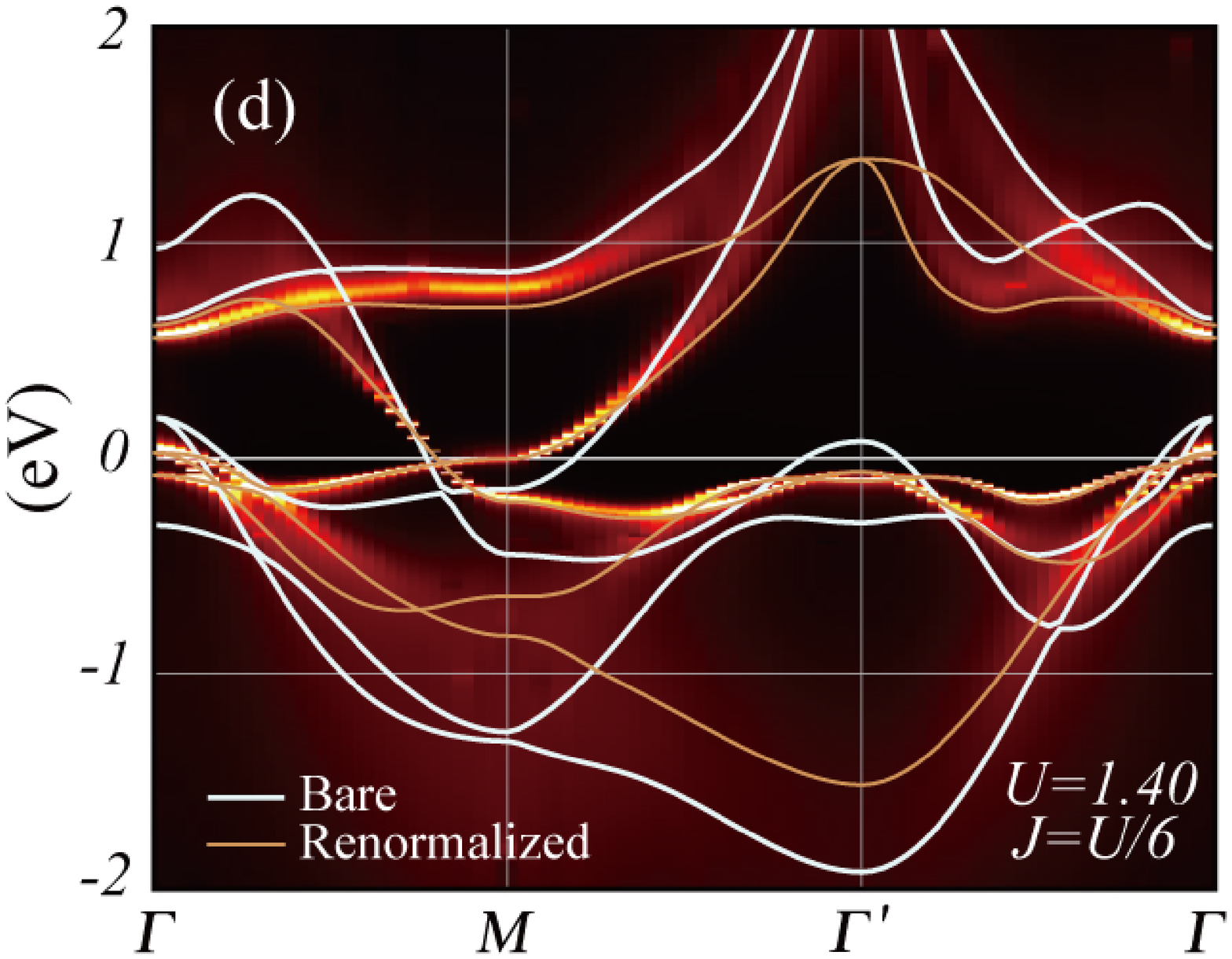}
\caption{(Color online) (a) Eigenvalue $\lambda$ as a function of $U$ in the FLEX approximation. (b) Spin susceptibility $\chi^s(\bm{q},0)$ at $U=1.40$ and (c) that at $U=1.80$ for $J=U/6$. (d) Spectral weight $\rho(\bm{k},\omega)$ along the high symmetry lines for $U=1.40$ and $J=U/6$. White and orange lines denote the unperturbed and the renormalized bands, respectively.}
\label{fig:5band}
\end{figure}

As mentioned in Sec. I, the FLEX calculations for the present model encounter severe problems as reported in Papers I and II.~\cite{rf:Ikeda,rf:Arita}
We start with summary and detailed analysis of the trouble points for the carrier density $n=6.00$ corresponding to the undoped parent compound LaFeAsO.
In Fig.~\ref{fig:5band}(a) we show the largest eigenvalue $\lambda$ of the Eliashberg equation as a function of $U$ for two choices of $J=U/6$ and $U/8$.
In both cases, the pairing symmetry for the maximum eigenvalue is a $s_\pm$-wave, as in the RPA calculations.
Although $\lambda$ initially increases with $U$, for $J=U/8$ it shows a tendency to saturate for $U\gtrsim 1.2$.~\cite{rf:diff}
For $J=U/8$, we cannot reach $\lambda=1$ even for large $U$, that is to say that even at $T=0.003\simeq 30$K superconductivity cannot be realized.
While we can obtain $\lambda=1$ for $J=U/6$ and $U\gtrsim 1.6$, the structure of the magnetic fluctuations is drastically changed, as discussed in Papers I and II.
With increasing $U$ the dominant magnetic fluctuation moves from the stripe-type AF with $\bm{Q}_1=(\pi,0)$ into a checkerboard-type AF with $\bm{Q}_2=(\pi,\pi)$, as illustrated in Figs.~\ref{fig:5band}(b) and \ref{fig:5band}(c).
This change is related to the renormalization of the band structure.

\begin{figure}
\centering
\vspace{10pt}
\includegraphics[width=70mm]{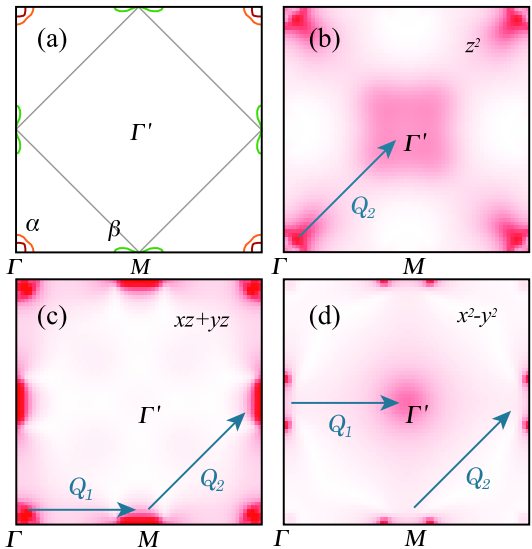}
\caption{(Color online) (a) The renormalized Fermi surface for $U=1.40$ and $J=U/6$. (b) The orbital-dependent weights of the Green's function $|\mathcal{G}_{\ell\ell}(\bm{k},i\pi T)|$ for $d_{3z^2-r^2}$-orbital, (c) $d_{yz}/d_{zx}$-orbital, and (d) $d_{x^2-y^2}$-orbital.}
\label{fig:qfs}
\end{figure}

In Fig.~\ref{fig:5band}(d), we show the spectral weight $\rho(\bm{k},\omega)=-\tfrac{1}{\pi}\sum_{\ell}{\rm Im}\mathcal{G}^R_{\ell\ell}(\bm{k},\omega)$,
obtained from the imaginary part of the retarded Green's function, and the quasiparticle band $\tilde{\xi}_{n\bm k}$ along the symmetry lines. 
To evaluate $\tilde{\xi}_{n\bm k}$ the self-energy is expanded to the first order in $\omega$, and the equation
\begin{equation}
\operatorname{Det}|(z_{\bm k}^{\ell}\omega-\mu)\delta_{\ell m}-
h^{\bm k}_{\ell m}-{\rm Re}\varSigma^{\rm R}_{\ell m}({\bm k},0)|=0 
\end{equation}
is solved. The mass enhancement factor $z_{\bm k}^{\ell}$ for each orbital, which is given by
\begin{equation}
z_{\bm k}^{\ell}=1-\frac{\partial \varSigma^{\rm R}_{\ell\ell}(\bm{k},\omega)}{\partial \omega}\biggr|_{\omega \to 0},
\end{equation}
can be at sufficiently low temperatures replaced with the approximate form 
\begin{equation}
z_{\bm k}^{\ell}\simeq 1-\frac{{\rm Im}\varSigma_{\ell\ell}(\bm{k},i\pi T)}{\pi T},
\end{equation}
while ${\rm Re}\varSigma_{\ell m}(\bm{k},i\pi T)$ is taken for the real part of the retarded self-energy ${\rm Re}\varSigma^{\rm R}_{\ell m}(\bm{k},0)$.
Thus obtained quasiparticle band traces closely the position of the peak in the the spectral density [the bright portion in Fig.~\ref{fig:5band}(d)].
The quasiparticle bands are strongly renormalized, which results in a drastic change in the Fermi surface [see Fig.~\ref{fig:qfs}(a)].
The Fermi surfaces around $\varGamma$ and $M$ shrink, and the $\gamma$ sheet around $\varGamma'$ vanishes completely.

To understand the effect of this drastic change on the dominant magnetic fluctuations, we show the orbital-dependent 
weight of the Green's function $|\mathcal{G}_{\ell\ell}(\bm{k},i\pi T)|$ in Figs.~\ref{fig:qfs}(b)$-$\ref{fig:qfs}(d).
Shown in Fig.~\ref{fig:qfs}(c), $d_{yz}/d_{zx}$ bands give rise to two nesting vectors, $\bm{Q}_1=(\pi,0)$ and $\bm{Q}_2=(\pi,\pi)$, as in the non-interacting case.
The renormalization of the  $d_{x^2-y^2}$ band leads to relative suppression of the $\bm{Q}_1$ fluctuation due to the reduction in the weight around $\varGamma'$.
On the other hand, the rise of the $d_{3z^2-r^2}$ band toward the Fermi level results in $\bm{Q}_2$ fluctuations.
Altogether, the band renormalization leads to 
strong enhancement of the $\bm{Q}_2$ fluctuation, which then dominates over the  $\bm{Q}_1$ fluctuation.

Up to now, however, there is no experimental evidence that $\bm{Q}_2$ fluctuation is strong
and that the renormalized bands drastically deviate from the LDA bands except for effective mass enhancement.
Even though, quite recently, the Fermi surface shrinkage has been reported in the vicinity of the AF critical point in As-P system~\cite{rf:Shishido}
and thus the tendency observed with the FLEX approximation seems to be right,~\cite{rf:Ortenzi} the extent of
the Fermi surface renormalization is overestimated. This leads us to conclude that the observed changes in the Fermi 
surface, caused by the shifts of the quasiparticle bands, are largely artifact of the LDA+FLEX approximation.
In particular, the deviation connected with the $d_{3z^2-r^2}$ orbital is serious.
In Paper II, we constructed and studied an effective four-band model excluding the $d_{3z^2-r^2}$ orbital.
However, even in the absence of $d_{3z^2-r^2}$, $\bm{Q}_1$ fluctuation is quite suppressed and $\bm{Q}_2$ channel becomes dominant in the intermediate correlation regime.
Thus, the straightforward application of FLEX for these models has the problem that the system cannot achieve the stripe-type AF phase transition.
In the following we will argue that the Fermi surface should not be substantially changed by the correlations and present a simple
way to achieve this within the present computational scheme.

\subsection{Modified self-energy and density relaxation}\label{sec:trouble}
We start with the empirical observation that LDA is surprisingly successful
in predicting the Fermi surface geometries even in complicated multi-band materials
with strong electronic correlations such as heavy fermion systems.~\cite{rf:Elgazzar,rf:Onuki}
This fact is even more striking when we realize that other quantities such as the effective
electron mass, spin susceptibility, or specific heat may be completely wrong.
Quite likely, this success of LDA is connected to its high accuracy in computing the
charge distributions, obtained by minimizing the density functional, which contains the
large electrostatic (Hartree) contribution. In real material, the large
Hartree term is the main restoring force which stabilizes the charge distribution.

Constructing the effective Hubbard model only the on-site interaction within the Fe $d$ shell is treated
explicitly while the other (large) interaction terms are absorbed into the fixed effective site energies
and hopping integrals, which do not depend on the charge distribution. Thus an important feedback
mechanism, which stabilizes the charge distribution is missing. 
For example, in the calculations of the previous section the renormalized 
$d_{3z^2-r^2}$ and $d_{x^2-y^2}$ occupancies deviate about $10\%$ from the unperturbed state.
This is remarkably large deviation since the Fermi surface is small in this system.
This leads to a remarkable modification of the small sheets of the Fermi surface.
An obvious solution to the problem of the missing feedback is a self-consistent recalculation of the effective Hamiltonian for the each FLEX iteration.
However, it is not feasible with our present computer codes.
Therefore we take an alternative ``poor man's'' approach. Taking LDA Fermi surface for realistic,
we restrict its modification due to correlations by subtracting the static part from the single-particle self-energy.
Namely, we replace the FLEX self-energy $\varSigma_{\ell m}(\bm{k},i\omega_n)$ in Eq.\eqref{eq:Gk} with
\begin{equation}
\delta\varSigma_{\ell m}(\bm{k},i\omega_n)
=\varSigma_{\ell m}(\bm{k},i\omega_n)-{\rm Re}\varSigma^{\rm R}_{\ell m}(\bm{k},0),
\end{equation}
where $\varSigma_{\ell m}(\bm{k},i\omega_n)$ is calculated in Eq.\eqref{eq:Sig} as usual, and  
${\rm Re}\varSigma^{\rm R}_{\ell m}(\bm{k},0)$ is its static energy part obtained by analytic continuation to the real axis of them.
Although a numerical analytic continuation generally includes errors, the $\omega \to 0$ limit at low temperatures can be obtained with high precisions
by the Pad\'e approximation using several lowest Matsubara frequencies.

\begin{figure}
\centering
\vspace{5pt}
\includegraphics[width=85mm]{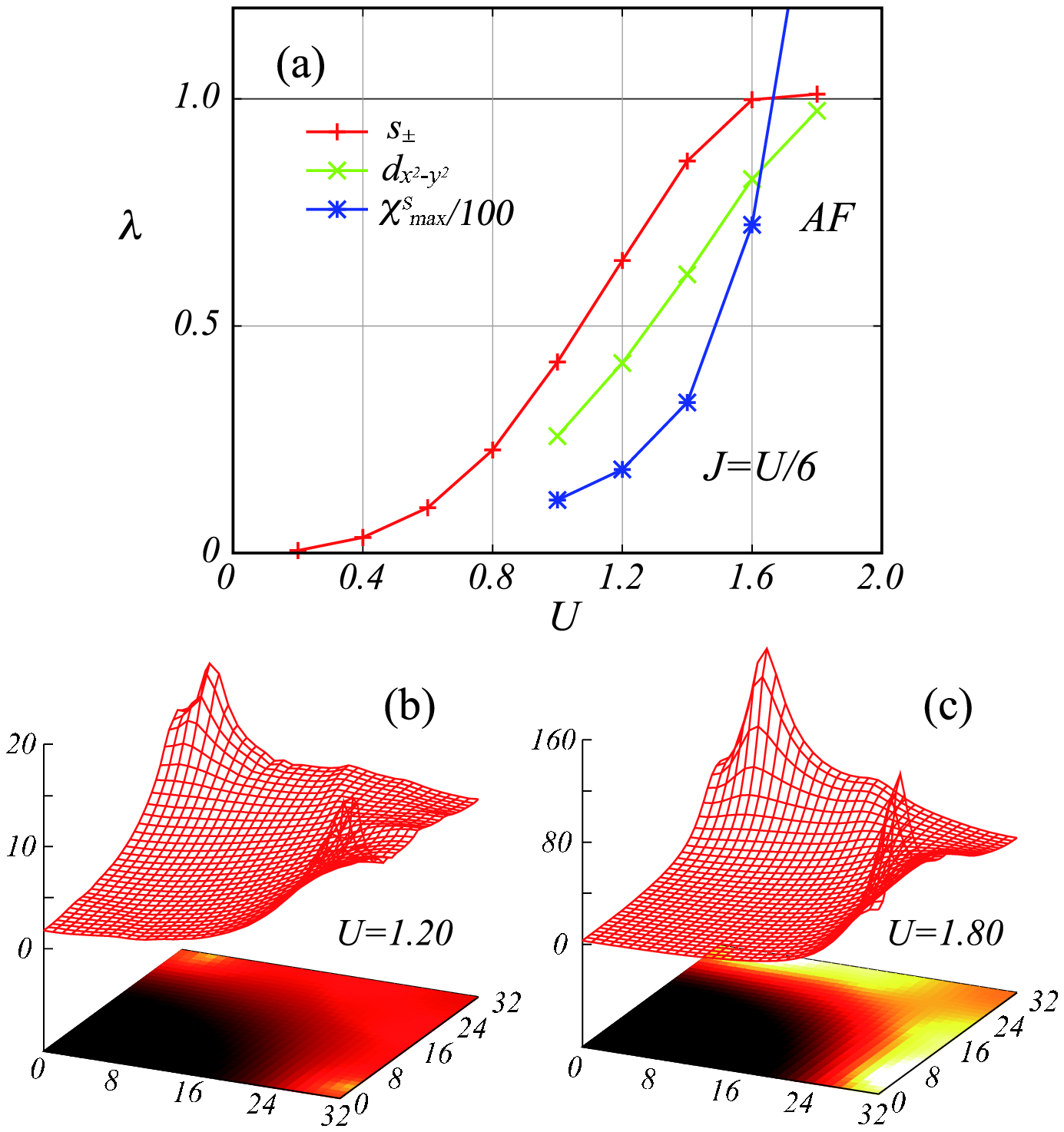}

\vspace{10pt}
\hspace{-4pt}\includegraphics[width=60mm]{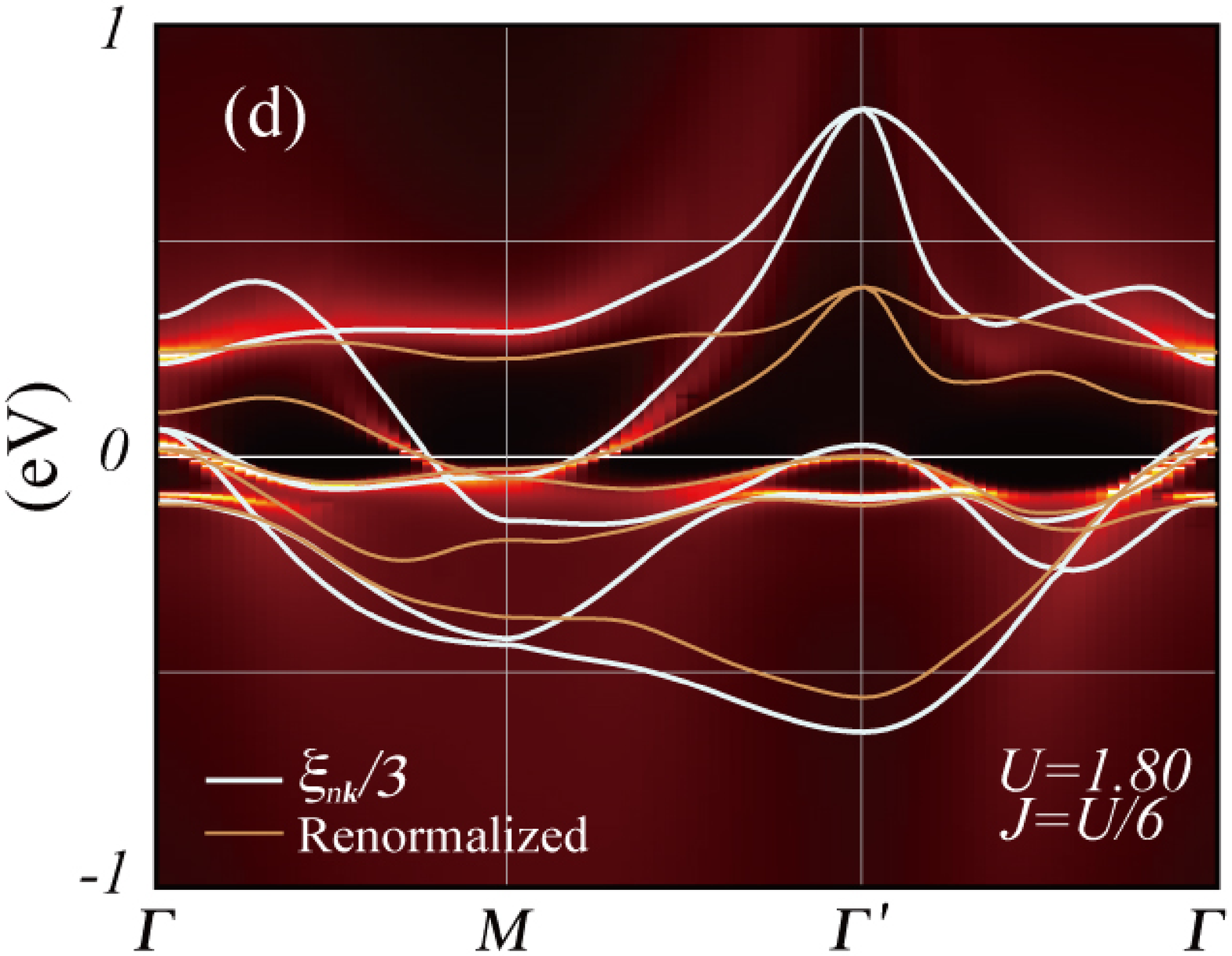}
\caption{(Color online) (a) $U$ dependence of the eigenvalue $\lambda$ and the maximum of the spin susceptibility $\chi^s(\bm{q},0)$ in the FLEX with modified self-energy. (b) Spin susceptibility $\chi^s(\bm{q},0)$ at $U=1.20$, and (c) that at $U=1.80$ for $J=U/6$. (d) Spectral weight $\rho(\bm{k},\omega)$ along the high symmetry lines for $U=1.80$ and $J=U/6$. White and orange lines denote the unperturbed band structure rescaled by $1/3$ and the renormalized band structure, respectively.}
\label{fig:mflex}
\end{figure}

\begin{figure}
\centering
\includegraphics[width=60mm]{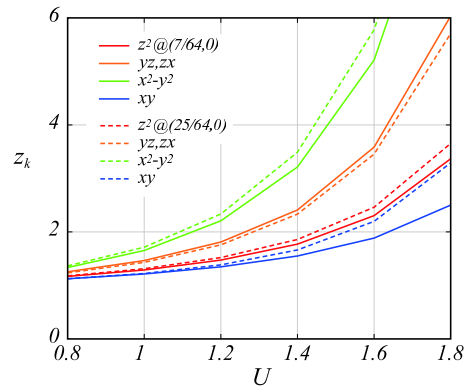}
\caption{(Color online) $U$ dependence of the mass enhancement factor for each orbital $z_{\bm k}^\ell$ at $(k_x,k_y)=(7\pi/32,0)$ and $(25\pi/32,0)$ with the same parameters as in Fig.~\ref{fig:mflex}(a).}
\label{fig:zk}
\end{figure}
\subsection{FLEX with modified self-energy}\label{sec:mFLEX}
Here, we present the LDA+FLEX results for the undoped model obtained with the modified self-energy.
In Fig.~\ref{fig:mflex}(a), we show the largest eigenvalue $\lambda$ and the maximum value of the spin susceptibility $\chi^s(\bm{q},0)$ as a function of $U$ for $J=U/6$.
The pairing symmetry corresponding to $\lambda$ remains $s_\pm$-wave as in \ref{sec:cFLEX}.
Although the value of $\lambda$ is almost the same as that in Fig.~\ref{fig:5band}(a) for $U \lesssim 1.0$, we can obtain a monotonic 
behavior also for the intermediate correlation regime, which is different from the erratic behavior in Fig.~\ref{fig:5band}(a). 
(where $\lambda$ is sensitive to a small change in parameters.)
This is because the electronic structure does not change drastically in this method as expected.
In fact, Fig.~\ref{fig:mflex}(b) and \ref{fig:mflex}(c) indicate that the structure of the spin fluctuation does not change even for $U=1.8$, and $(\pi,0)$ fluctuation is robust.
Strictly speaking, the peak position is not commensurate.
This drawback of the perfect elimination of the self-energy shift has no serious effect in comparison with the drastic change from $\bm{Q}_1$ to $\bm{Q}_2$ spin fluctuations.
In Fig.~\ref{fig:mflex}(d), we show the spectral weight $\rho(\bm{k},\omega)$ along the symmetry line for $U=1.8$.
The total bandwidth is reduced by about $1/3$ and the renormalization effect is larger near the Fermi level, but the Fermi-surface topology does not change.
In Fig.~\ref{fig:zk}, we present the mass enhancement factors for each orbital close to the $\alpha$ and $\beta$ surfaces.
The quasiparticle mass increases toward the AF critical point, which is a behavior recently
observed in the dHvA experiment for the As-P system~\cite{rf:Shishido}.

\section{Doping and Superconducting Properties}\label{sec:SC}
In the rest of the paper, we discuss the results obtained with the modified FLEX for various dopings
of the five-band model.
\subsection{Fermi surface}
\begin{figure}
\centering
\vspace{5pt}
\includegraphics[width=75mm]{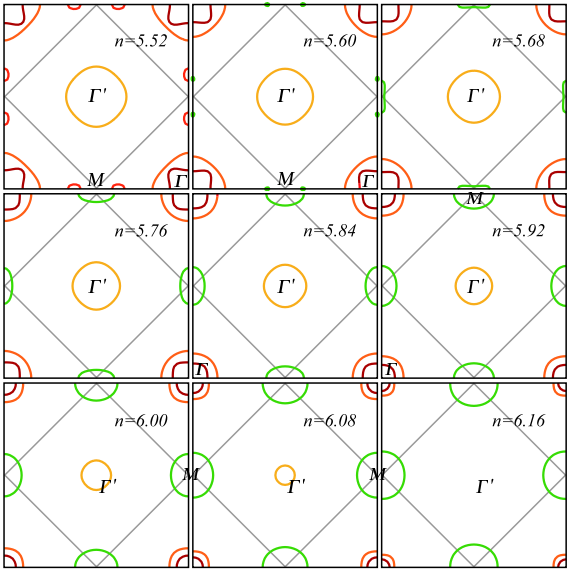}
\caption{(Color online) Evolution of the Fermi surface with the carrier doping. Green lines represent the electron surface, and the other reddish lines denote the hole surface. The electron surface around the $M$ points shrinks into a pair of Dirac points at $n=5.60$.}
\label{fig:fsevol}
\end{figure}

We start with the evolution of the Fermi surface with carrier doping.
Figure \ref{fig:fsevol} shows the Fermi surfaces of the five-band model for electron densities in the interval from 5.52 to 6.16, where
$n$=6.0 corresponds to the undoped LaFeAsO.
With electron doping, the $\gamma$ surface around $\varGamma'$ point shrinks and vanishes at $n\simeq 6.12$.
The disappearance of this sheet was shown to be important for the pseudogap behavior,~\cite{rf:Ikeda}
while its presence is important for the $(\pi,0)$ spin fluctuations.~\cite{rf:Ikeda,rf:Arita,rf:Kuroki2}
The electron $\beta$ surface around the $M$ point shrinks with hole doping.
At $n=5.60$, it reduces to a pair of Dirac points, and with further hole doping it becomes a hole surface as in $n=5.52$ case.
The same evolution is expected in (Ba,K)Fe$_2$As$_2$.
The Fermi surface at around $n=5.52$ is similar to that of the end material KFe$_2$As$_2$.~\cite{rf:TSato2}
Thus, the systematic calculation for doping dependence is relevant for the overall features of the phase diagrams and the gap symmetry in the related materials.

\subsection{Phase diagram}
\begin{figure}
\centering
\includegraphics[width=60mm]{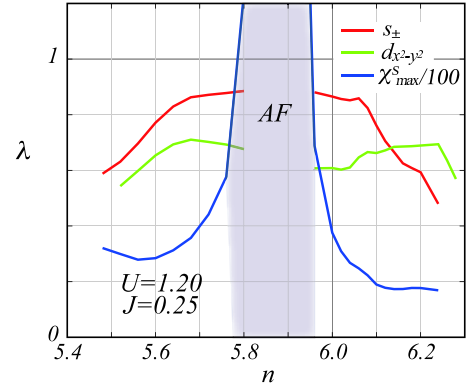}
\caption{(Color online) The phase diagram as a function of the carrier doping $n$ for $U=1.20$ and $J=0.25$. Red and green lines denote the eigenvalue $\lambda$ for the $s_\pm$-wave and $d_{x^2-y^2}$-wave, respectively. The blue line represents the maximum of the spin susceptibility $\chi^s(\bm{q},0)$. The shading denotes the AF phase.}
\label{fig:phase}
\end{figure}

Now, let us investigate the doping dependence of the largest eigenvalue $\lambda$ of the Eliashberg equation.
In Fig.~\ref{fig:phase}, we show the eigenvalue for $s_\pm$-wave and $d_{x^2-y^2}$-wave states, and the maximum value of the spin susceptibility $\chi^s(\bm{q},0)$ for $U=1.20$ and $J=0.25$.
There is a region of strong AF spin fluctuation on the hole-doped side ($n<6.00$) of the phase diagram.
The $s_\pm$-wave state overall dominates in the proximity of the AF phase.
The value of  $\lambda \sim 0.9$ suggests that the superconducting phase can be reached for strong interaction and/or lower temperature.
The eigenvalue $\lambda$ is remarkably reduced above $n \simeq 6.10$ and below $5.60$.
This corresponds to vanishing of the $\gamma$ hole surface and the $\beta$ electron surface at these points, respectively.
The $d_{x^2-y^2}$-wave solution dominates for $n \gtrsim 6.16$, where the $\gamma$ hole surface is absent.
The $s_\pm$-wave in this region develops nodes, so-called nodal $s_\pm$-wave~\cite{rf:Kuroki2},  as discussed in section \ref{sec:Gap}. 

\begin{figure}
\centering
\includegraphics[width=60mm]{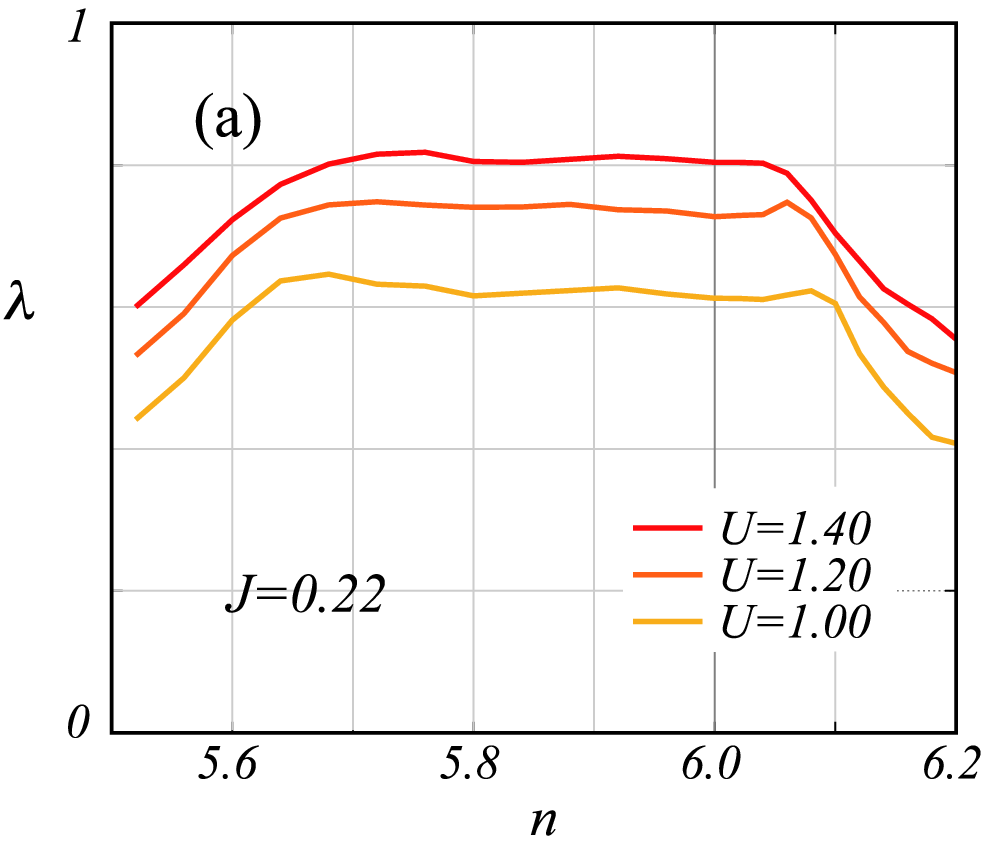}

\vspace{5pt}
\includegraphics[width=60mm]{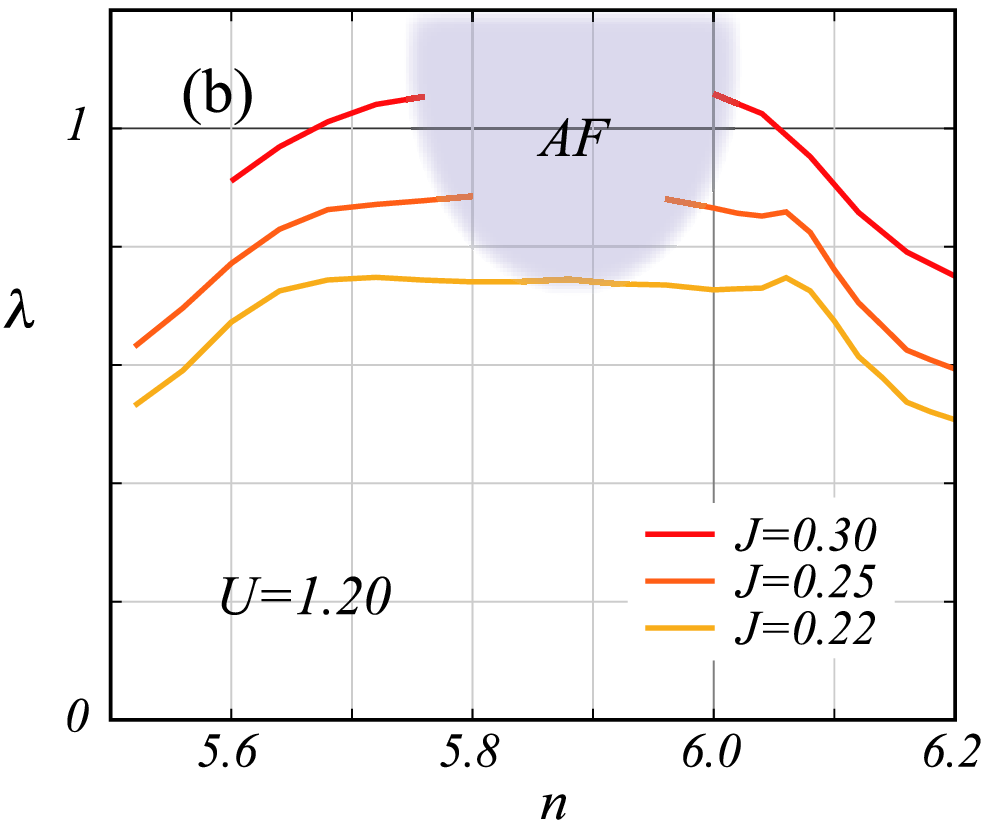}
\caption{(Color online) Doping dependence of the $s_\pm$-wave eigenvalue (a) for $U=1.40$, $1.20$, $1.00$ at $J=0.22$ and (b) for $J=0.30$, $0.25$, $0.22$ at $U=1.20$.}
\label{fig:phase2}
\end{figure}
Next, we examine the doping dependence of $s_\pm$-wave eigenvalue for several different parameters.
Figure \ref{fig:phase2}(a) shows the $\lambda$~vs~$n$ doping dependence for various $U$ at fixed $J=0.22$, and Fig.~\ref{fig:phase2}(b) for various $J$ at fixed $U=1.20$.
It is intriguing that $\lambda$ is almost flat over a rather wide doping range, and sensitive to $J$ rather than $U$.
Increase in $J$ enhances the AF spin fluctuation with $\bm{Q}=(\pi,0)$ and the eigenvalue $\lambda$.
In the large $J$ case, $\lambda$ increases when approaching the AF phase boundary, revealing a strong correlation between the AF spin fluctuation and 
superconducting $T_c$.
Two different behaviors of $\lambda$ can be distinguished,
(i) for relatively small $J$, small $\lambda$ insensitive to carrier doping, and
(ii) for relatively large $J$, large $\lambda$ sensitive to carrier doping and the presence of the AF phase.
These facts are consistent with the doping dependence of the transition temperature in LaFeAs(O,F) and (Ba,K)Fe$_2$As$_2$/Ba(Fe,Co)$_2$As$_2$, respectively.~\cite{rf:Ishida}

Next, to understand why SmFeAsO and NdFeAsO have the highest $T_c$ among the iron-pnictides, 
let us investigate the relationship between the pnictogen height $z$ and $T_c$,
which has been found experimentally,~\cite{rf:Lee} and stressed theoretically.~\cite{rf:Kuroki2}
We calculate the $s_\pm$-wave eigenvalue $\lambda$ for several different values of $z$, shown in Table~\ref{table1}, and the lattice parameters of LaFeAsO,
repeating the LDA+FLEX procedure described above for the respective crystal structures. (see Appendix \ref{sec:hopping})
\begin{table}
\caption{Height and internal $z$ coordinate of the As atom. La is shorthand of LaFeAsO. Nd and P 100\% represents LaFeAsO with the pnictogen height corresponding to NdFeAsO and LaFePO, respectively. The pnictogen height of P\! 50\% is an interpolated value between La and P\! 100\%.}\label{table1}
\begin{tabular}{lcc}
\hline\hline
  & pnictogen height (\AA) & internal coordinate $z$ \\ \hline
Nd & 1.38 & 0.6580 \\
La & 1.32 & 0.6512 \\
P\! 50\% & 1.23 & 0.6408 \\
P\! 100\% & 1.14 & 0.6304 \\ \hline\hline
\end{tabular}
\end{table}
\begin{figure}
\centering
\includegraphics[width=60mm]{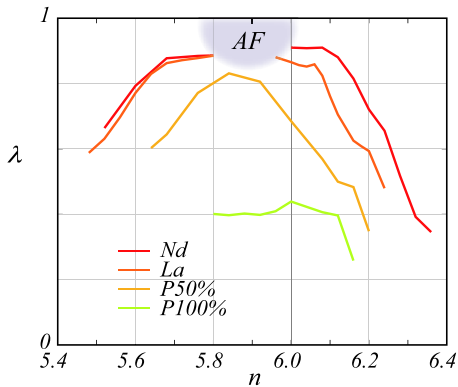}
\caption{(Color online) Doping dependence of the $s_\pm$-wave eigenvalue for several pnictogen heights with $U=1.20$ and $J=0.25$. The band structures for shorthand Nd, La, P $50\%$, and P $100\%$ have been obtained with the use of the pnictogen heights in Table \ref{table1}.}
\label{fig:height}
\end{figure}
\begin{figure}
\includegraphics[width=85mm]{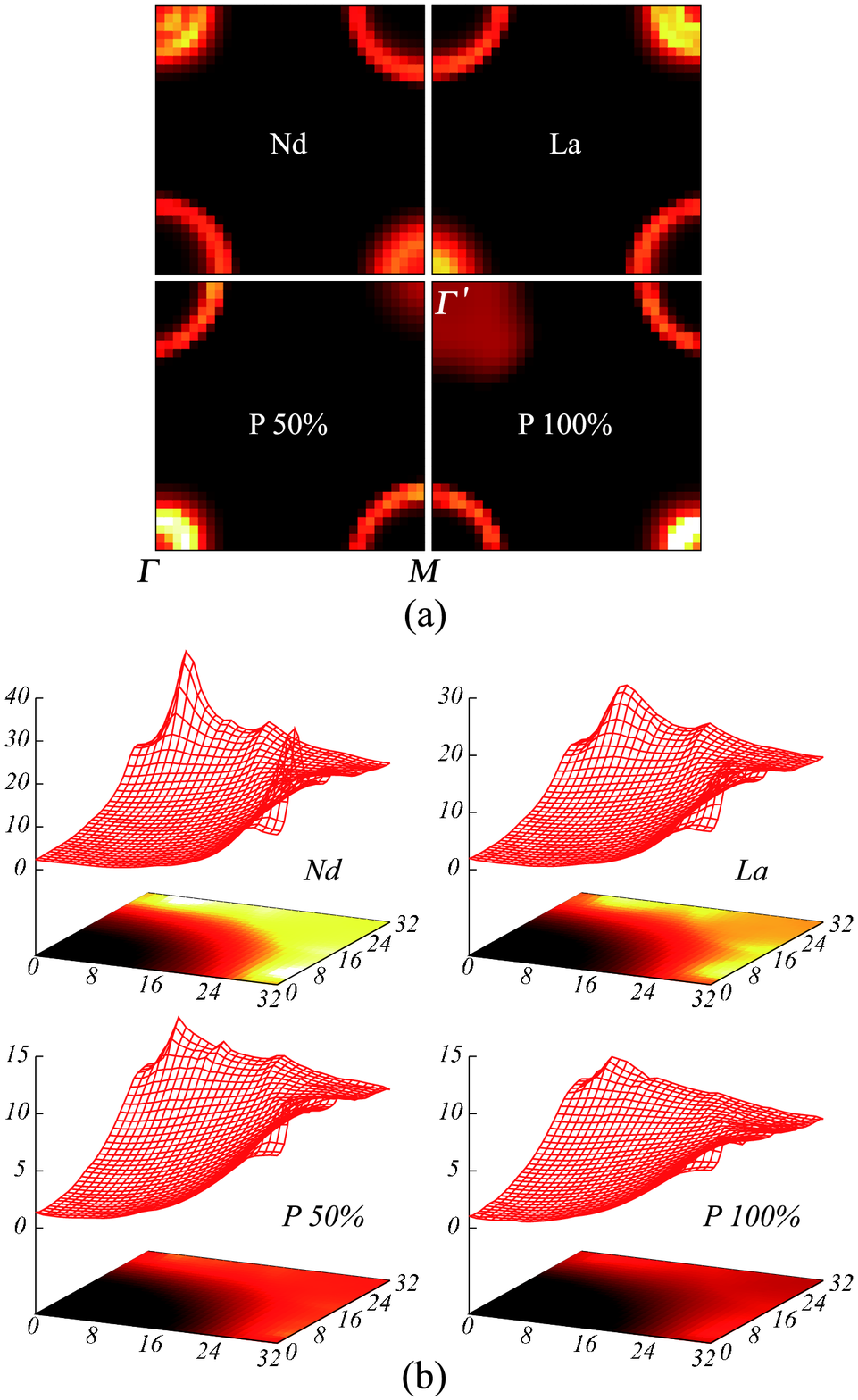}
\caption{(Color online) (a) Contour plot of the total weight of the Green's function $\sum_\ell |\mathcal{G}_{\ell\ell}(\bm{k},i\pi T)|$ and (b) Spin susceptibility $\chi^s(\bm{q},0)$ at $n=6.08$ for several pnictogen heights.}
\label{fig:n608h}
\end{figure}
In Fig.~\ref{fig:height}, we show the results for $U=1.20$ and $J=0.25$.
We can see that $\lambda$ grows with an increasing pnictogen height.
In the Nd case, $\lambda$ is larger for wider region, especially on the electron-doped side, than that in the La case.
Figures \ref{fig:n608h}(a) and \ref{fig:n608h}(b) show the total weight of the Green's function $\sum_\ell |\mathcal{G}_{\ell\ell}(\bm{k},i\pi T)|$ and the spin susceptibility $\chi^s(\bm{q},0)$ at $n=6.08$.
The bright parts in the former represent the Fermi surface.
With the decreasing pnictogen height, the $\gamma$ hole surface around $\varGamma'$ point visibly shrinks and becomes dim.
Correspondingly, the $(\pi,0)$ spin fluctuation is suppressed.
Thus, the pnictogen height is very important for the high $T_c$ since it controls
 the size of the $\gamma$ hole surface and the magnitude of the spin fluctuation.
The observed trend is consistent with the experimental data showing that NdFeAsO and SmFeAsO hold high $T_c$ even for heavily electron doping.

\subsection{Gap anisotropy}\label{sec:Gap}
In this section, we examine anisotropy of the gap functions for various dopings. 
To this end, we plot the band diagonal anomalous Green's function,
\begin{equation}
\mathcal{F}_{nn}(\bm{k},i\pi T)=
\sum_{\ell m}\big(u^{\bm k}_{\ell n}\big)^*\mathcal{F}_{\ell m}(\bm{k},i\pi T)u^{\bm k}_{mn},
\end{equation}
obtained from Eq.\eqref{eq:Fk} with the eigenvectors of the Eliashberg Eq.~\eqref{eq:elia}.
It singles out the gap amplitude on the Fermi surface.

\begin{figure}
\centering
\includegraphics[width=85mm]{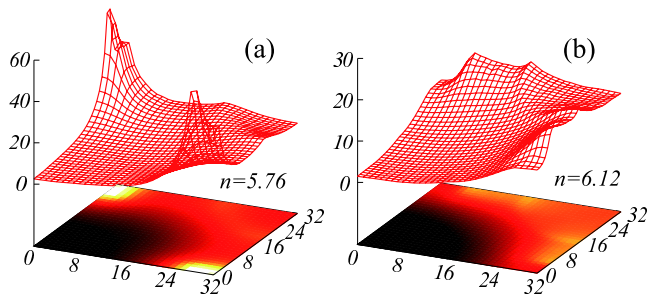}
\caption{(Color online) Spin susceptibility $\chi^s(\bm{q},0)$ (a) at $n=5.76$ and (b) at $n=6.12$ for $U=1.20$ and $J=0.25$.}
\label{fig:chR}
\end{figure}

\begin{figure}
\centering
\vspace{5pt}
\includegraphics[width=80mm]{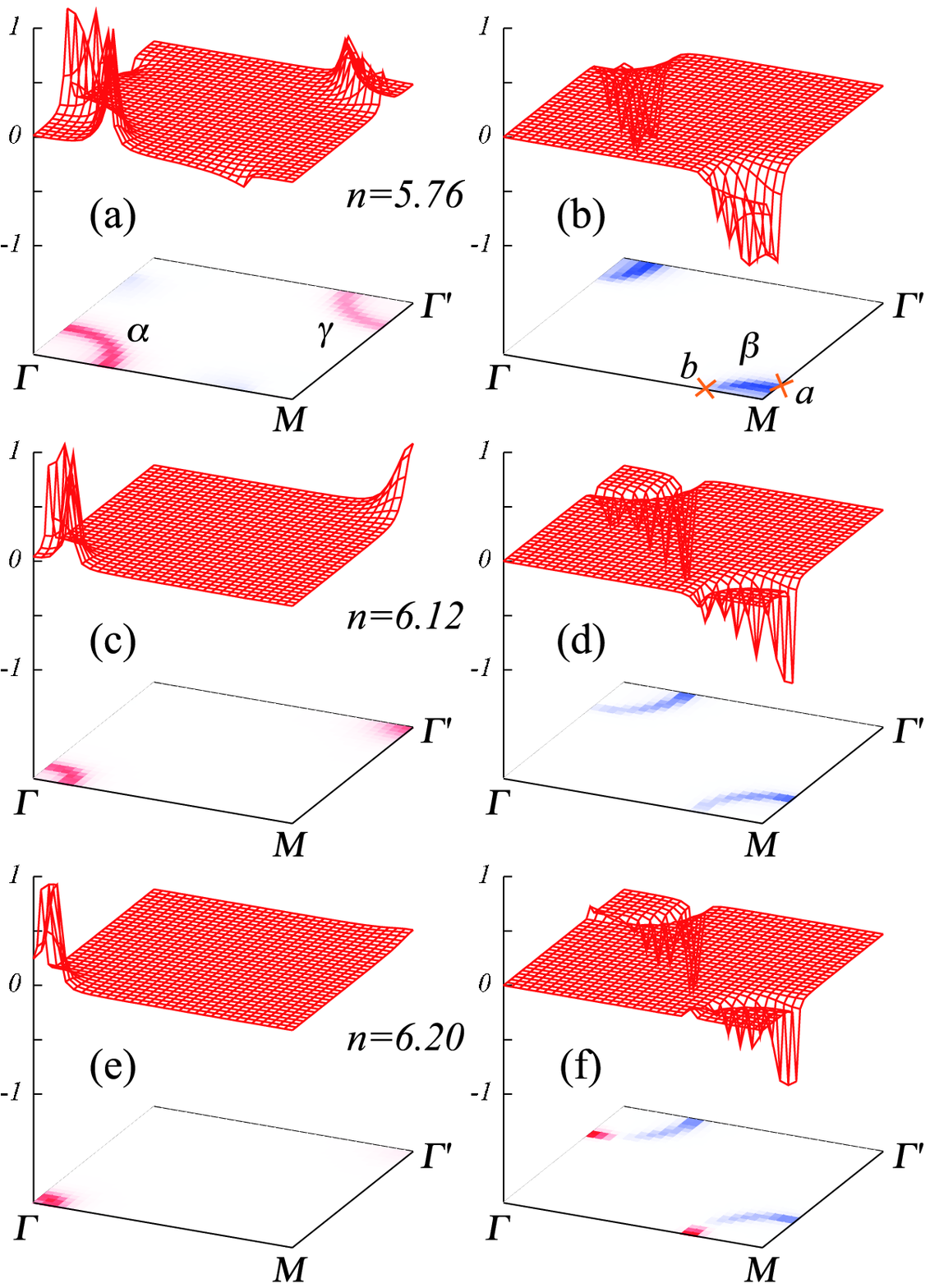}
\caption{(Color online) Band-diagonal anomalous Green's function $\mathcal{F}_{nn}(\bm{k},i\pi T) $ for the $s_\pm$-wave at $n=5.76$, $6.12$ and $6.20$. (a), (c), (e) correspond to the anomalous Green's function on the third band, and (b), (d), (f) on the second band~\cite{rf:another}.}
\label{fig:Gap}

\centering
\vspace{10pt}
\includegraphics[width=80mm]{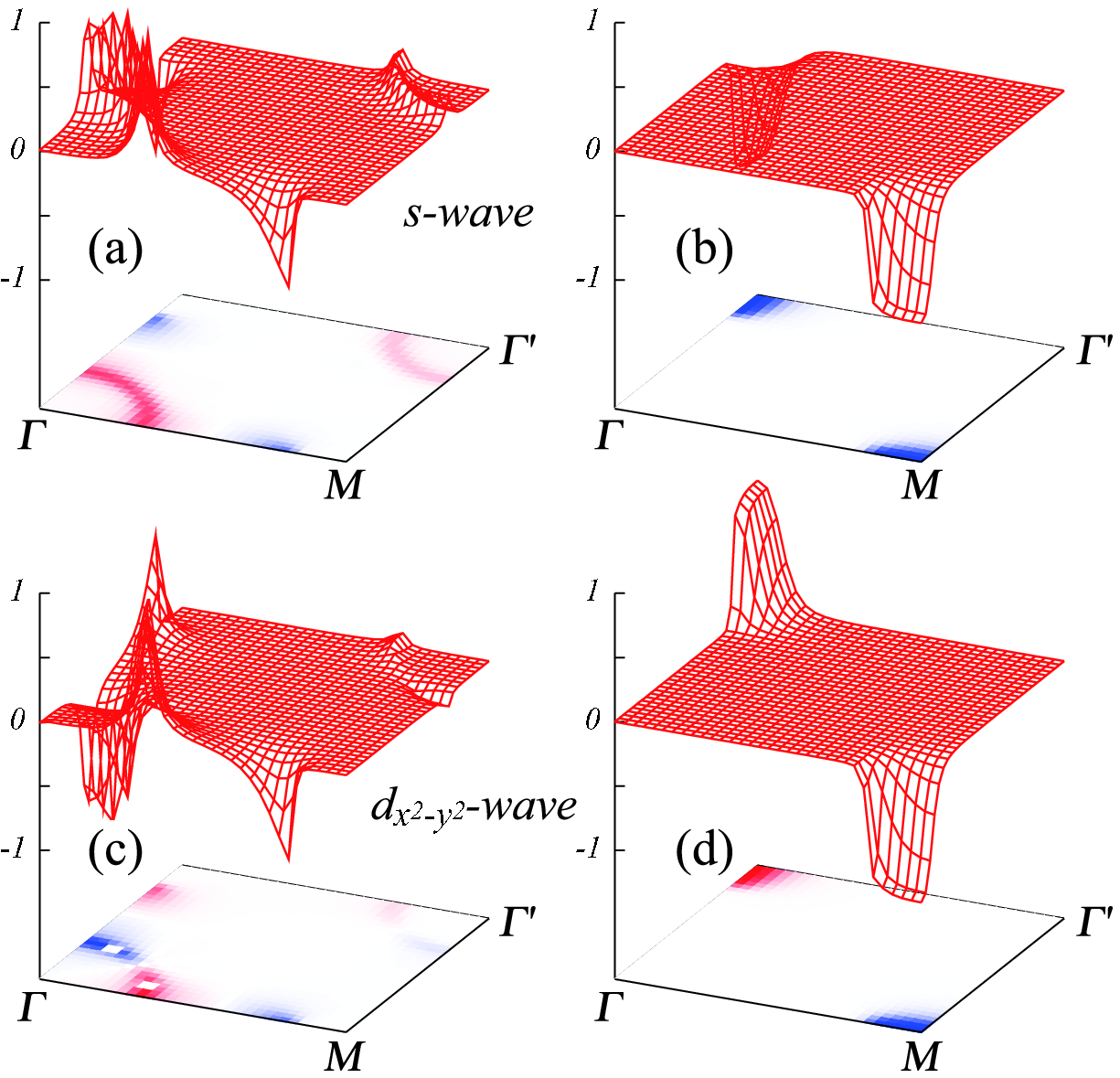}
\caption{(Color online) Band-diagonal anomalous Green's function $\mathcal{F}_{nn}(\bm{k},i\pi T) $ at $n=5.60$. (a) and (b) correspond to the third and the second bands in the $s_\pm$-wave case, respectively. (c) and (d) in the $d_{x^2-y^2}$-wave case.}
\label{fig:n560}
\end{figure}

In the hole-doped region, the strong $(\pi,0)$ spin fluctuations render the $s_\pm$-wave a likely gap function.
For instance, at $n=5.76$, the $(\pi,0)$ spin fluctuation is remarkably enhanced as shown in Fig.~\ref{fig:chR}(a), 
and the gap function has no nodes as shown in Figs.~\ref{fig:Gap}(a) and \ref{fig:Gap}(b).
In this case, the gap function on the $\alpha$ surface has almost the same amplitude as that on the $\beta$ surface 
(with opposite sign), and about a half of the $\gamma$ surface one.
This is consistent with the gap structure observed in ARPES on (Ba,K)Fe$_2$As$_2$.~\cite{rf:Ding,rf:Nakayama}
The ratio of the $\alpha$ to $\gamma$ surface gap amplitude changes gradually with carrier doping, 
as illustrate by $n=5.60$ in Fig.~\ref{fig:n560}(a).

In the electron-doped region, the AF spin fluctuations are suppressed as shown in Fig.~\ref{fig:chR}(b).
While the $s_\pm$-wave  pairing symmetry is still favorable, the gap function becomes remarkably anisotropic 
on the $\beta$ surface, as shown in Fig.~\ref{fig:Gap}(d).
Further electron doping leads to a sign reversal, i.e., it becomes so-called nodal $s_\pm$-wave shown in Fig.~\ref{fig:Gap}(f).
However, the corresponding eigenvalue is small, and the $d_{x^2-y^2}$-wave pairing prevails.
Thus we may expect a fully-gapped state in the hole-doped region, and gap minima or line-nodes in the electron-doped region.
This may be the key to understanding of the material-dependent nodal features as the fully-gapped behavior in (Ba,K)Fe$_2$As$_2$, 
and the nodal/nodeless behaviors in Ba(Fe,Co)$_2$As$_2$ and LaFeAs(O,F).

Let us take a detailed look at the gap anisotropy.
Even at $n=5.76$, the gap amplitude around the $M$ point exhibits some anisotropy.
Comparing the gap amplitudes at the points marked $a$ and $b$ in Fig.~\ref{fig:Gap}(b), we find that the amplitude at $a$ is the same as that on the $\alpha$ surface, while the amplitude at $b$ is about the same as on the $\gamma$ surface.
This reflects the fact that the Fermi surface at $a$ is dominated by the $d_{yz}/d_{zx}$ orbitals, 
and that at $b$  by the $d_{x^2-y^2}$ orbital.
Thus, the gap amplitude on the $\beta$ surface will be always anisotropic.
The appearance of the distinct gap minima on the $\beta$ surface in the electron doped region 
is related to the shrinkage of the $\gamma$ surface.

Finally, let us comment on  the end material of the 122 series, KFe$_2$As$_2$.
Quit recently, it has been reported that it exhibits a two-gap nodal behavior.~\cite{rf:Fukazawa2,rf:Dong}
In our model, this material corresponds to a hole doping of 0.5, close to the filling $n=5.60$, which leads to a Dirac-cone band structure.
Since the gap functions for these two fillings barely differ, let us discuss only the $n=5.60$ case,
in which the unrenormalized $\beta$ surface reduces to a point.
In the strong coupling theory of superconductivity, electrons within a finite energy window around the Fermi level 
participate in the pairing, and the anomalous Green's function can have a large amplitude even at $\bm k$ points far from the Fermi surface.
Indeed, we find a large weight in a finite region around the $M$ point, see Figs.~\ref{fig:n560}(b) and \ref{fig:n560}(d).
Another interesting point is the small magnitude of the gap function on the $\gamma$ surface relative
to that around the $\varGamma$ and $M$ points,  both in the $s_\pm$-wave and $d_{x^2-y^2}$-wave states.
Observation of two gap of distinct sizes is consistent with the recent experiments in KFe$_2$As$_2$.~\cite{rf:Fukazawa2}
Nevertheless, this conclusion must be taken with care.
In addition to the small value of $\lambda$, we overestimate the $d_{3z^2-r^2}$-orbital contribution in the heavily hole-doped region.
In the present two-dimensional model, the $c$-axis dispersion due to the $d_{3z^2-r^2}$ orbital is neglected.
Although it is not crucial as long as the $d_{3z^2-r^2}$ band is far from the Fermi level, 
its presence close to the Fermi level suppresses the eigenvalue of the $d_{x^2-y^2}$-wave relative to the $s_\pm$-wave.
Therefore, we expect the $d_{x^2-y^2}$-wave eigenvalue to be comparable with the $s_\pm$-wave one in a more realistic three-dimensional calculation.

\subsection{Pairing Glue}
\begin{figure}
\centering
\includegraphics[width=60mm]{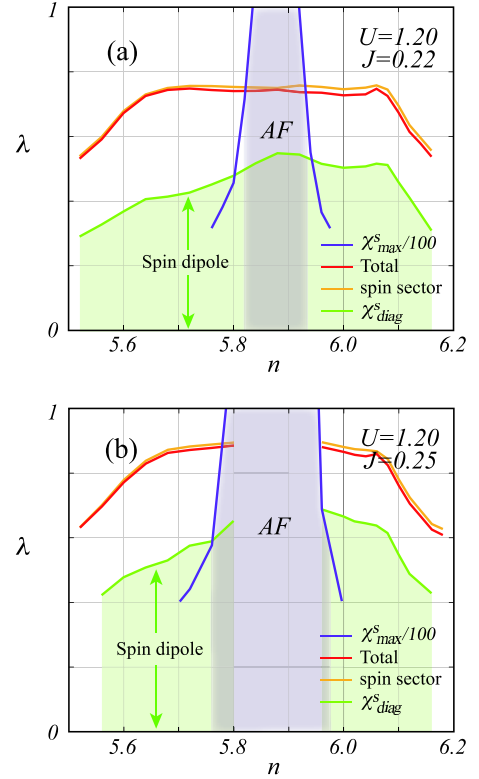}
\caption{(Color online) Phase diagram for (a) $J=0.22$ and (b) $J=0.25$ at $U=1.20$. The violet line represents the maximum of the spin susceptibility $\chi^s(\bm{q},0)$. The red, orange, and green lines denote the eigenvalue $\lambda$ with the total contribution, the spin sector and just the conventional AF spin fluctuation, respectively.}
\label{fig:mech}
\end{figure}

In this section, we analyze the paring mechanism.
Although we have obtained the phase diagram insensitive to carrier doping and the presence of the AF phase in relatively small $J$ case, we need to clarify what is the glue for the superconducting pairs in this case.
In Fig.~\ref{fig:mech}, we investigate the $J=0.22$ and $0.25$ cases.
Within the RPA and FLEX formalism the pairing interaction can be separated into 
contributions from the spin and charge sectors.
The former includes all spin fluctuations described by  $\chi^s_{12,34}$, and the latter consists of all charge/orbital fluctuations contained in $\chi^c_{12,34}$.
In Fig.~\ref{fig:mech}, the ``spin sector'' denotes contribution of $\chi^s_{12,34}$ only,
i.e., we omit $\chi^c_{12,34}$ when solving the Eliashberg equation. Clearly, the contribution of the charge sector is negligible.

Next, we consider only the contribution from the orbital-diagonal spin fluctuations $\chi^s_{\ell\ell,mm}$.
These add up to the spin susceptibility  $\chi^s_{\rm diag}\equiv \langle\!\langle\bm{S}^\mu_i,\bm{S}^\mu_j\rangle\!\rangle=\sum_{\ell m}\chi^s_{\ell\ell,mm}$,
which describes the correlations between spin dipole moments $\bm{S}^\mu_i=\sum_{\ell,\alpha\beta}c^\dagger_{i\ell\alpha}\sigma^\mu_{\alpha\beta}c_{i\ell\beta}$.
In Fig.~\ref{fig:mech}, we denote this contribution with ``spin dipole.''
We find that for both choices of $J$, the contribution of the spin dipole-dipole fluctuations reflects the proximity of the AF phase, that is, the eigenvalue $\lambda$ overall increases towards the AF phase boundary.
Such a behavior is rather remarkable in the light of flatness of the full $\lambda$ vs $n$.
Moreover, the dipole-only eigenvalue is only $60-70\%$ of the total value of $\lambda$.
The rest comes from the orbital-off-diagonal spin fluctuations, which represent correlations
between higher-order multipoles, especially, $\chi^s_{\ell\ell,\ell m}+\chi^s_{\ell\ell,m\ell}$ as suggested in Paper I,
This gives a coupling between spin-dipole and spin-quadrupole.
Thus, although the conventional AF spin fluctuations provide the largest contribution to the paring glue, 
the higher-order multipolar fluctuations given by the off-diagonal elements of $\chi^s_{12,34}$ assist the superconducting pairing, and push the transition temperature up.
Therefore, the correlation between $T_c$ and the proximity of the AF phase is seemingly weak.

\section{Summary and Conclusions}
Using a combination of {\it ab initio} band structure and the FLEX approximation, we have constructed
a five-band model of iron pnictides and studied the doping dependence of superconductivity. 
We have employed a simple procedure to fix the shape of the Fermi surface to its LDA shape. 

We have found that the superconductivity is stable in a wider interval for hole doping than for electron doping.
The $s_\pm$-wave state corresponds to the largest eigenvalue of the Eliashberg equation $\lambda$ over a wide doping range, 
only in the heavily electron doped region $d$-wave state becomes more favorable.
For a relatively small Hund's coupling $J$, the eigenvalue $\lambda$ is relatively small and insensitive to carrier doping.
On the other hand, for large $J$, $\lambda$ is large and more sensitive to the carrier doping and the proximity of AF phase.
These observations are consistent with the behavior of the superconducting transition temperature $T_c$ in LaFeAs(O,F) and (Ba,K)Fe$_2$As$_2$/Ba(Fe,Co)$_2$As$_2$, respectively.
Furthermore, to understand why the highest $T_c$ is found in SmFeAsO and NdFeAsO, we have investigated the influence of the pnictogen height on $\lambda$.
The pnictogen height has a great impact on the size of the Fermi surface around the $\varGamma'$ point formed by the $d_{x^2-y^2}$ orbital.
In the Nd case, the $\varGamma'$ sheet is large resulting in a large $\lambda$ over a wider doping region, especially on the electron-doped side, than in the La system.
This agrees with the fact that SmFeAsO and NdFeAsO hold the high $T_c$ even when heavily electron doped.

As for the anisotropy of the gap function,  
we find a fully-gapped state in the hole-doped region, and remarkably anisotropic gap function around the $M$ point in the electron-doped region.
This may be the key to understanding of the material-dependent nodal behavior, such as fully-gapped behavior in (Ba,K)Fe$_2$As$_2$, and nodal/nodeless behavior in LaFeAs(O,F).
Furthermore, we have found indications of gaps of two distinct sizes for the end 122 material KFe$_2$As$_2$.

Finally, concerning the pairing mechanism, we have explained why the correlation between $T_c$ and the presence of the AF phase is seemingly weak in this system.
Only $60-70\%$ of the total pairing interaction originates from the diagonal components of spin fluctuation, which corresponds to the conventional spin-spin correlation.
The remaining part originates from correlations involving higher-order spin multipoles.
This additional pairing glue naturally comes from the multi-band character of iron-pnictides with several different
orbital contributions at the Fermi surface.
Therefore, it is an oversimplification to attribute pairing in these materials only to the conventional AF spin fluctuations.

\begin{acknowledgments}
We are grateful to Y. Matsuda, T. Shibauchi, K. Ishida, Y. Nakai, S. Kasahara, K. Hashimoto, H. Shishido, T. Takimoto, and K. Yamada for valuable discussions.
This work is supported by a Grant-in-Aid for Scientific Research on Priority Areas (Grant No. 20029014) and the Global COE Program ``The Next Generation of Physics, Spun from Universality and Emergence'' from the Ministry of Education, Culture, Sports, Science and Technology, Japan.
JK acknowledges the support of the SFB 484 of the Deutsche Forschungsgemeinschaft.
\end{acknowledgments}

\appendix
\section{Hopping integrals}\label{sec:hopping}
Let us here provide a set of in-plane hopping integrals on the MLWFs basis, $t_{\ell m}[\varDelta\bar x,\varDelta\bar y]$, for each band calculation in Table \ref{table1}, where $[\varDelta\bar x,\varDelta\bar y]$ denotes the in-plane hopping vector, and $(\ell m)$ the orbitals.
Note that the $\bar x$- and $\bar y$-axes point toward neighboring Fe atoms while $x$ and $y$ in the orbitals ($1:d_{3z^2-r^2}$, $2:d_{xz}$, $3:d_{yz}$, $4:d_{x^2-y^2}$, $5:d_{xy}$) are those in the coordinate system for the original unit cell.
In Table \ref{table2}, we list $t_{\ell m}[\varDelta\bar x,\varDelta\bar y]$ up to the fifth neighbors, the magnitude of which is larger than $0.005$eV.
We can obtain the principal hopping integrals for each band structure from this table and 
the relation $t_{\ell m}[\varDelta\bar x,\varDelta\bar y]=t_{m\ell}[-\varDelta\bar x,-\varDelta\bar y]$.

\begin{table*}
\caption{Hopping integrals $t_{\ell m}[\varDelta\bar x,\varDelta\bar y]$ for Nd, La, P\! 50\%, and P\! 100\% in Table \ref{table1} in units of eV. $\sigma_y$, $I$, and $\sigma_d$ mean $t_{\ell m}[\varDelta\bar x,-\varDelta\bar y]$, $t_{\ell m}[-\varDelta\bar x,-\varDelta\bar y]$, and $t_{\ell m}[\varDelta\bar y,\varDelta\bar x]$, respectively.  ``$\pm$'' and ``$\pm(\ell'm')$'' in the columns denote $\pm t_{\ell m}[\varDelta\bar x,\varDelta\bar y]$ and $\pm t_{\ell'm'}[\varDelta\bar x,\varDelta\bar y]$, respectively. }\label{table2}
\begin{tabular}{crrrrrrccccrrrrrrccc}
\hline\hline
\addlinespace[2pt]
&\multicolumn{9}{c}{\hspace{-10pt} (Nd)~~~~ $[\varDelta\bar x,\varDelta\bar y]$}
&\makebox[1pt]
&\multicolumn{9}{c}{\hspace{-10pt} (La)~~~~ $[\varDelta\bar x,\varDelta\bar y]$}\\[1pt]
\cline{2-10}\cline{12-20}
\addlinespace[2pt]
$(\ell m)$
&$[0,0]$&$[1,0]$&$[1,1]$&$[2,0]$&$[2,1]$&$[2,2]$&$\sigma_y$&$I$&$\sigma_d$ &
&$[0,0]$&$[1,0]$&$[1,1]$&$[2,0]$&$[2,1]$&$[2,2]$&$\sigma_y$&$I$&$\sigma_d$ \\[1pt]
\cline{1-10}\cline{12-20}
\addlinespace[1pt]
$(11)$&7.872&$-$0.033&$-$0.006&$-$0.025&     0.020&$-$0.011&$+$     &$+$&$+$ &&
7.949&$-$0.053&              &$-$0.029&     0.023&$-$0.011&$+$      &$+$&$+$ \\
$(12)$&         &$-$0.083&              &              &              &               &$-(13)$&$-$&$-$ &&
         &$-$0.075&              &              &              &              &$-(13)$&$-$&$-$ \\
$(13)$&         &     0.083&$-$0.153&              &              &$-$0.027&$-(12)$&$-$&$+$ &&
         &     0.075&$-$0.147&              &              &$-$0.028&$-(12)$&$-$&$+$ \\
$(14)$&         &              &     0.138&              &      0.007&$-$0.013&$-$      &$+$&$+$ &&
         &              &     0.160&              &     0.008&$-$0.014&$-$       &$+$&$+$ \\
$(15)$&         &$-$0.294&              &$-$0.008&$-$0.018&              &$+$      &$+$&$-$ &&
         &$-$0.298&              &              &$-$0.021&              &$+$      &$+$&$-$ \\
$(22)$&8.075&$-$0.189&     0.135&     0.005&     0.009&               &$+(33)$&$+$&$+$ &&
8.141&$-$0.201&     0.136&     0.006&     0.009&              &$+(33)$&$+$&$+$ \\
$(23)$&         &     0.130&              &     0.021&$-$0.017&               &$+$      &$+$&$-$ &&
         &     0.132&              &     0.022&$-$0.016&              &$+$       &$+$&$-$ \\
$(24)$&         &     0.168&              &              &     0.009&               &$+(34)$&$-$&$-$ &&
         &     0.169&              &              &     0.013&              &$+(34)$&$-$&$-$ \\
$(25)$&         &$-$0.235&     0.127&              &$-$0.007&     0.006&$-(35)$&$-$&$+$ &&
         &$-$0.250&     0.135&              &$-$0.008&     0.007&$-(35)$&$-$&$+$ \\
$(33)$&8.075&$-$0.189&     0.310&     0.005&$-$0.025&     0.061&$+(22)$&$+$&$+$ &&
8.141&$-$0.201&     0.327&     0.006&$-$0.026&     0.065&$+(22)$&$+$&$+$ \\
$(34)$&         &     0.168&     0.046&              &     0.018&               &$+(24)$&$-$&$+$ &&
         &     0.169&     0.023&              &     0.019&               &$+(24)$&$-$&$+$ \\
$(35)$&         &     0.235&              &              &     0.024&               &$-(25)$&$-$&$-$ &&
         &     0.250&              &              &     0.027&               &$-(25)$&$-$&$-$ \\
$(44)$&8.174&     0.121&     0.108&$-$0.019&$-$0.027&$-$0.024&$+$      &$+$&$+$ &&
8.288&     0.151&     0.119&$-$0.025&$-$0.030&$-$0.025&$+$       &$+$&$+$\\
$(45)$&         &              &              &              &$-$0.008&               &$-$      &$+$&$-$ &&
         &              &              &              &$-$0.010&               &$-$       &$+$&$-$ \\
$(55)$&7.761&     0.310&$-$0.058&$-$0.016&              &               &$+$     &$+$&$+$ &&
7.829&     0.315&$-$0.065&$-$0.019&              &               &$+$      &$+$&$+$ \\
\hline
\end{tabular}
\begin{tabular}{crrrrrrccccrrrrrrccc}
\hline
\addlinespace[2pt]
&\multicolumn{9}{c}{\hspace{-18pt} (P 50\%)~~~ $[\varDelta\bar x,\varDelta\bar y]$}
&\makebox[1pt]
&\multicolumn{9}{c}{\hspace{-20pt} (P 100\%)~~~ $[\varDelta\bar x,\varDelta\bar y]$}\\[1pt]
\cline{2-10}\cline{12-20}
\addlinespace[2pt]
$(\ell m)$
&$[0,0]$&$[1,0]$&$[1,1]$&$[2,0]$&$[2,1]$&$[2,2]$&$\sigma_y$&$I$&$\sigma_d$ &
&$[0,0]$&$[1,0]$&$[1,1]$&$[2,0]$&$[2,1]$&$[2,2]$&$\sigma_y$&$I$&$\sigma_d$ \\[1pt]
\cline{1-10}\cline{12-20}
\addlinespace[1pt]
$(11)$&8.055&$-$0.083&     0.015&$-$0.034&     0.028&$-$0.011&$+$      &$+$&$+$ &&
8.196&$-$0.108&     0.034&$-$0.036&     0.032&$-$0.012&$+$      &$+$&$+$ \\
$(12)$&         &$-$0.060&              &     0.008&$-$0.008&               &$-(13)$&$-$&$-$ &&
         &$-$0.041&              &     0.016&$-$0.012&               &$-(13)$&$-$&$-$ \\
$(13)$&         &     0.060&$-$0.133&$-$0.008&     0.008&$-$0.029&$-(12)$&$-$&$+$ &&
         &     0.041&$-$0.113&$-$0.016&     0.017&$-$0.030&$-(12)$&$-$&$+$ \\
$(14)$&         &              &     0.193&              &     0.007&$-$0.015&$-$       &$+$&$+$ &&
         &              &     0.225&              &     0.005&$-$0.014&$-$       &$+$&$+$ \\
$(15)$&         &$-$0.304&              &     0.005&$-$0.025&               &$+$      &$+$&$-$ &&
         &$-$0.306&              &     0.016&$-$0.029&              &$+$      &$+$&$-$ \\
$(22)$&8.231&$-$0.219&     0.135&     0.011&$-$0.009&     0.005&$+(33)$&$+$&$+$ &&
8.356&$-$0.232&     0.132&     0.017&$-$0.010&     0.006&$+(33)$&$+$&$+$ \\
$(23)$&         &     0.133&              &     0.024&$-$0.014&               &$+$       &$+$&$-$ &&
         &     0.128&              &     0.022&$-$0.010&              &$+$       &$+$&$-$ \\
$(24)$&         &     0.166&              &              &     0.019&               &$+(34)$&$-$&$-$ &&
         &     0.161&              &              &     0.025&              &$+(34)$&$-$&$-$ \\
$(25)$&         &$-$0.272&     0.147&              &$-$0.010&     0.009&$-(35)$&$-$&$+$ &&
         &$-$0.291&     0.157&     0.008&$-$0.013&     0.011&$-(35)$&$-$&$+$ \\
$(33)$&8.231&$-$0.219&     0.347&     0.011&$-$0.030&     0.071&$+(22)$&$+$&$+$ &&
8.356&$-$0.232&     0.356&     0.017&$-$0.035&     0.076&$+(22)$&$+$&$+$ \\
$(34)$&         &     0.166&$-$0.023&              &     0.020&                &$+(24)$&$-$&$+$ &&
         &     0.161&$-$0.075&              &     0.022&               &$+(24)$&$-$&$+$ \\
$(35)$&         &     0.272&              &              &     0.033&                &$-(25)$&$-$&$-$ &&
         &     0.291&              &$-$0.008&     0.039&               &$-(25)$&$-$&$-$ \\
$(44)$&8.454&     0.200&     0.134&$-$0.036&$-$0.035&$-$0.026&$+$      &$+$&$+$ &&
8.652&     0.245&     0.145&$-$0.049&$-$0.040&$-$0.026&$+$      &$+$&$+$ \\
$(45)$&         &              &              &              &$-$0.013&                &$-$       &$+$&$-$ &&
         &              &              &              &$-$0.017&               &$-$       &$+$&$-$ \\
$(55)$&7.911&     0.322&$-$0.077&$-$0.026&              &                &$+$      &$+$&$+$ &&
8.022&     0.328&$-$0.089&$-$0.034&     0.005&               &$+$      &$+$&$+$ \\
\hline\hline
\end{tabular}
\end{table*}

\section{Symmetry Consideration for the hopping matrix}\label{sec:symH}
We here give careful consideration to symmetry operations in the two-dimensional model Hamiltonian.
The unperturbed Hamiltonian is defined as follows:
\begin{subequations}
\label{eq:H0}
\begin{alignat}{2}
H_0&=&\sum_{\bm{k}\ell m \sigma}h^{\bm k}_{\ell m}c^\dag_{\bm{k}\ell\sigma}c_{\bm{k}m\sigma} \\
&=&\sum_{\bm{k}\ell m \sigma}\tilde{h}^{\bm k}_{\ell m}\tilde{c}^\dag_{\bm{k}\ell\sigma}\tilde{c}_{\bm{k}m\sigma} \\
&=&\sum_{\bm{k}n\sigma}\epsilon_{\bm{k}n}a^\dag_{\bm{k}n\sigma}a_{\bm{k}n\sigma},
\end{alignat}
\end{subequations}
where $c^\dag_{\bm{k}\ell\sigma}$, and $c_{\bm{k}m\sigma}$ are the Fourier transforms of 
$c^\dag_{i\ell\sigma}$, and $c_{im\sigma}$.
The hopping matrix $h^{\bm k}_{\ell m}$ has the following form:
\begin{equation}
h^{\bm k}_{\ell m}=
\begin{pmatrix}
\rm e & i\rm o & i\rm o & \rm e & \rm e \cr
-i\rm o & \rm e & \rm e & i\rm o & i\rm o \cr
-i\rm o & \rm e & \rm e & i\rm o & i\rm o \cr
\rm e & -i\rm o & -i\rm o & \rm e & \rm e \cr
\rm e & -i\rm o & -i\rm o & \rm e & \rm e
\end{pmatrix},
\end{equation}
where ``~$i$~'' is the imaginary unit, and ``~$\rm e$~'' and ``~$\rm o$~'' denote, respectively, even- and odd-parity functions of the wave vector $\bm{k}$.
Introducing a sign function with $s=+/-$ below/above the line $k_x=k_y$ in Fig.~\ref{fig:unfold}(b),
$i{\rm o}=is{\rm |o|}$ with ${\rm |o|}=s{\rm o}$, which is unchanged under the inversion.
In this case, using the transformation
\begin{equation}
\label{eq:ctilde}
c^\dag_{\bm{k}\ell\sigma} = \bm{i}_{\bm{k}\ell}^{\ *}\tilde{c}^\dag_{\bm{k}\ell\sigma}, ~~~~
c_{\bm{k}m\sigma} = \bm{i}_{\bm{k}m}\tilde{c}_{\bm{k}m\sigma}
\end{equation}
with 
\begin{equation}
\label{eq:im}
\bm{i}_{\bm{k}m}=
\begin{pmatrix}
1 &&&& \cr
& is &&& \cr
&& is && \cr
&&& 1& \cr
&&&& 1
\end{pmatrix},
\end{equation}
$h^{\bm k}_{\ell m}$ can be cast into a real symmetric form, 
\begin{equation}
\label{eq:hk}
\tilde{h}^{\bm k}_{\ell m}=\bm{i}_{\bm{k}\ell}^{\ *} h^{\bm k}_{\ell m} \bm{i}_{\bm{k}m}=
\begin{pmatrix}
\rm e & \rm -|o| & \rm -|o| & \rm e & \rm e \cr
\rm -|o| & \rm e & \rm e & \rm |o| & \rm |o| \cr
\rm -|o| & \rm e & \rm e & \rm |o| & \rm |o| \cr
\rm e & \rm |o| & \rm |o| & \rm e & \rm e \cr
\rm e & \rm |o| & \rm |o| & \rm e & \rm e
\end{pmatrix},
\end{equation}
which is unchanged under the inversion.
The third line of Eq.\eqref{eq:H0} shows the diagonal form obtained from $\tilde{h}^{\bm k}_{\ell m}$ by the unitary transformation $\tilde{u}_{\ell n}^{\bm k}$.
The band indices $n$ are set according to the main orbital component, not in the order of energy eigenvalues. 
In this case, the unitary matrix $\tilde{u}_{mn}^{\bm k}$ has the same matrix form as the above $\tilde{h}^{\bm k}_{\ell m}$, 
\begin{equation}
\tilde{u}_{mn}^{\bm k}=
\begin{pmatrix}
\rm e & \rm |o| & \rm |o| & \rm e & \rm e \cr
\rm |o| & \rm e & \rm e & \rm |o| & \rm |o| \cr
\rm |o| & \rm e & \rm e & \rm |o| & \rm |o| \cr
\rm e & \rm |o| & \rm |o| & \rm e & \rm e \cr
\rm e & \rm |o| & \rm |o| & \rm e & \rm e
\end{pmatrix}.
\end{equation}
With the use of this unitary matrix, the direct transformation from $c_{\bm{k}m\sigma}$ to $a_{\bm{k}n\sigma}$ can be defined by
\begin{subequations}
\begin{alignat}{2}
c_{\bm{k}m\sigma} &=~u^{\bm k}_{mn}a_{\bm{k}n\sigma} 
&&=\bm{i}_{\bm{k}m}\tilde{u}_{mn}^{\bm k}\bm{i}_{\bm{k}n}^{\ *} a_{\bm{k}n\sigma} \\[3pt]
c^\dag_{\bm{k}\ell\sigma}~ &= \big(u^{\bm k}_{\ell n}\big)^* a^\dag_{\bm{k}n\sigma}
&&= \bm{i}_{\bm{k}\ell}^{\ *}\tilde{u}^{\bm k}_{\ell n} \bm{i}_{\bm{k}n} a^\dag_{\bm{k}n\sigma}
\end{alignat}
\end{subequations}
In this case, the relation 
\begin{equation}
\big( u_{mn}^{\bm k} \big)^*=u_{mn}^{-\bm{k}},
\end{equation}
corresponding to the time-reversal symmetry of the Hamiltonian naturally holds.
Thus, we can fix the complicated phase factors accompanying the diagonalization at each $\bm{k}$ point.
In addition, considering the mirror symmetry about the two lines $k_x=k_y$ and $k_x+k_y=2\pi$ through the $\varGamma'$ point, 
we can carry out any numerical calculations in the reduced zone as shown in Fig.~\ref{fig:unfold}(b),
which improves the accuracy and speeds up the calculation.

\section{Green's functions}\label{sec:Green}
We here summarize several generic relations for the Green's functions.
First, let us define the normal and anomalous Green's functions for orbitals $\ell$ and $m$.
We assume a paramagnetic normal state, and the spin-singlet symmetry for the superconducting state.
Then the Green's functions at the wave vector $\bm k$ and the imaginary time $\tau$ are given by
\begin{subequations}
\begin{align}
\mathcal{G}_{\ell m}(\bm{k},\tau)
&=\langle\!\langle c_{\bm{k}\ell\sigma}(\tau)c^\dag_{\bm{k}m\sigma}(0) \rangle\!\rangle, \\[3pt]
\sigma\mathcal{F}_{\ell m}(\bm{k},\tau)
&=\langle\!\langle c_{\bm{k}\ell\sigma}(\tau)c_{\bm{-k}m\bar\sigma}(0) \rangle\!\rangle, \\[2pt]
\sigma\mathcal{F}^\dag_{\ell m}(\bm{k},\tau)
&=\langle\!\langle c^\dag_{\bm{-k}\ell\bar\sigma}(\tau)c^\dag_{\bm{k}m\sigma}(0) \rangle\!\rangle.
\end{align}
\end{subequations}
Here,
$\langle\!\langle A(\tau)B(0) \rangle\!\rangle = -\langle T_\tau [A(\tau)B(0)] \rangle$
with the conventional notation.
From the above definition and the time-reversal invariance,
we obtain the following relations:
\begin{subequations}
\begin{alignat}{3}
\mathcal{G}_{m\ell}(\bm{k},\tau)^*~
&=~\mathcal{G}_{\ell m}(\bm{k},\tau)
&&=\mathcal{G}_{\ell m}(-\bm{k},\tau)^*,\\[3pt]
\mathcal{F}_{m\ell}(-\bm{k},-\tau)
&=~\mathcal{F}_{\ell m}(\bm{k},\tau)
&&=\mathcal{F}_{\ell m}(-\bm{k},\tau)^*, \\[3pt]
=\mathcal{F}^\dag_{m\ell}(-\bm{k},\tau)
&=\mathcal{F}^\dag_{\ell m}(\bm{k},-\tau)
&&=\mathcal{F}^\dag_{\ell m}(-\bm{k},-\tau)^*.
\end{alignat}
\end{subequations}
By the Fourier transformation from $\bm k$ to $\bm r$, these relations are rewritten as follows,
\begin{subequations}
\begin{alignat}{3}
\label{eq:Gr}
\mathcal{G}_{m\ell}(-\bm{r},\tau)^*
&=~\mathcal{G}_{\ell m}(\bm{r},\tau) 
&&=\mathcal{G}_{\ell m}(\bm{r},\tau)^*, \\[3pt]
\mathcal{F}_{m\ell}(-\bm{r},-\tau)
&=~\mathcal{F}_{\ell m}(\bm{r},\tau)
&&=\mathcal{F}_{\ell m}(\bm{r},\tau)^*, \\[3pt]
=\mathcal{F}^\dag_{m\ell}(-\bm{r},\tau)~
&=\mathcal{F}^\dag_{\ell m}(\bm{r},-\tau)
&&=\mathcal{F}^\dag_{\ell m}(\bm{r},-\tau)^*.
\end{alignat}
\end{subequations}
Thus, $\mathcal{G}_{\ell m}(\bm{r},\tau)$ and $\mathcal{F}^{(\dag)}_{\ell m}(\bm{r},\tau)$ are real functions.
On the other hand, by the Fourier transformation from $\tau$ to the fermionic Matsubara frequency $\omega_n=(2n+1)\pi T$, we can obtain 
\begin{subequations}
\begin{eqnarray}
\label{eq:Gw}
&&\hspace{-5pt}\mathcal{G}_{m\ell}(\bm{k},-i\omega_n)^*
=\mathcal{G}_{\ell m}(\bm{k},i\omega_n) 
=\mathcal{G}_{\ell m}(-\bm{k},-i\omega_n)^*, \hspace{30pt} \\[3pt]
&&\hspace{-7pt}\mathcal{F}_{m\ell}(-\bm{k},-i\omega_n)
=\mathcal{F}_{\ell m}(\bm{k},i\omega_n)
=\mathcal{F}_{\ell m}(-\bm{k},-i\omega_n)^*, \\[3pt]
&&\hspace{-10pt}=\mathcal{F}^\dag_{m\ell}(-\bm{k},i\omega_n)
=\mathcal{F}^\dag_{\ell m}(\bm{k},-i\omega_n)
=\mathcal{F}^\dag_{\ell m}(-\bm{k},i\omega_n)^*.
\end{eqnarray}
\end{subequations}
Concerning the wave vector $\bm k$, each component of the normal Green's functions holds the same irreducible representation in the space group as the corresponding hopping matrix $h^{\bm k}_{\ell m}$.
Depending on the parity even or odd, the last equality in Eq.\eqref{eq:Gw} becomes
\begin{equation}
\mathcal{G}_{\ell m}(\bm{k},i\omega_n) = \pm\mathcal{G}_{\ell m}(\bm{k},-i\omega_n)^*.
\end{equation}
By Eqs.\eqref{eq:ctilde} and \eqref{eq:im}, it is convenient to introduce Green's functions,
\begin{subequations}
\begin{align}
\tilde{\mathcal{G}}_{\ell m}(\bm{k},i\omega_n) &=\!\int_0^\beta d\tau e^{i\omega_n\tau}
\langle\!\langle \tilde{c}_{\bm{k}\ell\sigma}(\tau)\tilde{c}^\dag_{\bm{k}m\sigma} \rangle\!\rangle, \\
\tilde{\mathcal{F}}_{\ell m}(\bm{k},i\omega_n) &=\!\int_0^\beta d\tau e^{i\omega_n\tau}
\langle\!\langle \tilde{c}_{\bm{k}\ell\sigma}(\tau)\tilde{c}_{-\bm{k}m\bar\sigma} \rangle\!\rangle,
\end{align}
\end{subequations}
in $\tilde{c}_{\bm{k}m\sigma}$ representation with the inverse temperature $\beta=1/T$.
In this case, 
\begin{subequations}
\begin{align}
\mathcal{G}_{\ell m}(\bm{k},i\omega_n)
&=\bm{i}_{\bm{k}\ell} \ \tilde{\mathcal{G}}_{\ell m}(\bm{k},i\omega_n) \ \bm{i}_{\bm{k}m}^{\ *}, \\[2pt]
\mathcal{F}_{\ell m}(\bm{k},i\omega_n)
&=\bm{i}_{\bm{k}\ell} \ \tilde{\mathcal{F}}_{\ell m}(\bm{k},i\omega_n) \ \bm{i}_{\bm{k}m}^{\ *},
\end{align}
\end{subequations}
with $\bm{i}_{-\bm{k}m}=\bm{i}_{\bm{k}m}^{\ *}$, and then,
\begin{subequations}
\begin{alignat}{3}
\tilde{\mathcal{G}}_{\ell m}(\bm{k},i\omega_n)
&=\tilde{\mathcal{G}}_{\ell m}(\bm{k},-i\omega_n)^*
&&=\tilde{\mathcal{G}}_{m\ell}(\bm{k},i\omega_n), \\[2pt]
\tilde{\mathcal{F}}_{\ell m}(\bm{k},i\omega_n)
&=\tilde{\mathcal{F}}_{\ell m}(\bm{k},-i\omega_n)^*
&&=\tilde{\mathcal{F}}_{m\ell}(\bm{k},i\omega_n)^*.
\end{alignat}
\end{subequations}
Thus, we can eliminate the sign depending on the parity.
In addition to these relations, a value at $\bm k$ point in Green's function holds a simple relation to that at the star of $\bm{k}$ under the space-group symmetry.
Thus, Green's functions at all $\bm k$ points in the unfolded BZ can be generated from those values at $\bm k$ points in the reduced zone.
These relations obtained in this section are practical in actual calculations.

\section{FLEX formalism}\label{sec:FLEX}
We here summarize the formulation of FLEX, following Ref.\cite{rf:Takimoto}.
The linearized Dyson-Gorkov equations for spin-singlet pairing with the abbreviation $k=(\bm{k},i\omega_n)$ are given by
\begin{subequations}
\begin{align}
\label{eq:Gk}\mathcal{G}_{\ell m}(k) &= \mathcal{G}_{\ell m}^0(k)
+\mathcal{G}_{\ell\ell'}^0(k)\varSigma_{\ell'm'}(k)\mathcal{G}_{m'm}(k), \\[3pt]
\mathcal{F}_{\ell m}(k) &= \mathcal{G}_{\ell\ell'}(k)\mathcal{G}_{mm'}(-k)\varDelta_{\ell'm'}(k) \\[3pt]
\label{eq:Fk}&=\mathcal{G}_{\ell\ell'}(k)\mathcal{G}_{mm'}(k)^*\varDelta_{\ell'm'}(k),
\end{align}
\end{subequations}
where $\varSigma_{\ell m}(k)$ and $\varDelta_{\ell m}(k)$ are the normal and the anomalous self-energies, respectively.
In $\tilde{c}_{\bm{k}m\sigma}$ representation, these equations can be rewritten as
\begin{subequations}
\begin{align}
\label{eq:Gkt}\tilde{\mathcal{G}}_{\ell m}(k) &= \tilde{\mathcal{G}}_{\ell m}^0(k)
+\tilde{\mathcal{G}}_{\ell\ell'}^0(k)\tilde{\varSigma}_{\ell'm'}(k)\tilde{\mathcal{G}}_{m'm}(k), \\[3pt]
\label{eq:Fkt}\tilde{\mathcal{F}}_{\ell m}(k) &= \tilde{\mathcal{G}}_{\ell\ell'}(k)\tilde{\mathcal{G}}_{mm'}(k)^*\tilde{\varDelta}_{\ell'm'}(k),
\end{align}
\end{subequations}
where $\tilde{\varSigma}_{\ell m}(k)$ and $\tilde{\varDelta}_{\ell m}(k)$ are, respectively, defined by
\begin{subequations}
\begin{align}
\label{eq:Sigt}\tilde{\varSigma}_{\ell m}(k) &= \bm{i}_{\bm{k}\ell}^{\ *} \varSigma_{\ell m}(k) \bm{i}_{\bm{k}m}, \\[3pt]
\label{eq:Dkt}\tilde{\varDelta}_{\ell m}(k) &= \bm{i}_{\bm{k}\ell}^{\ *} \varDelta_{\ell m}(k) \bm{i}_{\bm{k}m}.
\end{align}
\end{subequations}
In the FLEX approximation, the self-energies are described with effective interactions composed of ladder and bubble diagrams as follows,
\begin{subequations}
\begin{align}
\label{eq:Sig}\varSigma_{\ell m}(k) &= \sum_q V_{\ell\ell',mm'}(q)\mathcal{G}_{\ell'm'}(k-q), \\
\label{eq:Dk}\varDelta_{\ell m}(k) &= -\sum_q V^s_{\ell\ell',m'm}(q)\mathcal{F}_{\ell'm'}(k-q),
\end{align}
\end{subequations}
where the effective interactions $V_{\ell\ell',mm'}(q)$ and $V^s_{\ell\ell',mm'}(q)$ are, respectively, given by $(\ell\ell', mm')$ element of matrices, $\hat{V}(q)$ and $\hat{V}^s(q)$.
Here, $\nu_n=2n\pi T$ in $q=(\bm{q},i\nu_n)$ is the bosonic Matsubara frequency.
In the present study, for simplicity, let us consider only particle-hole processes and omit particle-particle ladder processes in these effective interactions.
This simplification can be expected to be justified not only qualitatively but also semi-quantitatively
in the case where the spin fluctuation dominates.
In this case, the effective interactions are, in the matrix form, given by 
\begin{subequations}
\begin{align}
\label{eq:Vq}
\begin{array}{l}
\displaystyle
\phantom{\frac{3}{2}}\hat{V}(q)~=\hat{U}_{\uparrow\downarrow}-2\hat{U}_{\uparrow\uparrow}-\hat{U}_{\uparrow\downarrow}\hat{\chi}_0(q)\hat{U}_{\uparrow\downarrow}, \\
\displaystyle
\hspace{64pt}+\frac{3}{2}\hat{U}^s\hat{\chi}^s(q)\hat{U}^s+\frac{1}{2}\hat{U}^c\hat{\chi}^c(q)\hat{U}^c \\
\end{array}
\\
\label{eq:Vs}
\hat{V}^s(q) =\hat{U}_{\uparrow\downarrow}+\frac{3}{2}\hat{U}^s\hat{\chi}^s(q)\hat{U}^s-\frac{1}{2}\hat{U}^c\hat{\chi}^c(q)\hat{U}^c,
\end{align}
\end{subequations}
with the bare vertices $\hat{U}^{s,c}=\hat{U}_{\uparrow\downarrow}\mp\hat{U}_{\uparrow\uparrow}$ and three susceptibilities, $\hat{\chi}_0(q)$, $\hat{\chi}^s(q)$ and $\hat{\chi}^c(q)$.
Each component of the bare vertices is defined by
\begin{subequations}
\begin{align}
(\hat{U}_{\uparrow\downarrow})_{\ell\ell,\ell\ell}~~ &=U \\
(\hat{U}_{\uparrow\downarrow})_{\ell\ell,mm} &=U' \\
(\hat{U}_{\uparrow\downarrow})_{\ell m,\ell m} &=J \\
(\hat{U}_{\uparrow\downarrow})_{\ell m,m \ell} &=J'
\end{align}
\end{subequations}
\begin{subequations}
\begin{align}
(\hat{U}_{\uparrow\uparrow})_{\ell\ell,mm} &=U'-J \\
(\hat{U}_{\uparrow\uparrow})_{\ell m,\ell m} &=J-U'
\end{align}
\end{subequations}
where $\ell \ne m$ and the other components are zero.

Susceptibilities $\hat{\chi}^s(q)$ and $\hat{\chi}^c(q)$ represent, respectively, susceptibilities for spin sector and charge sector, which include the RPA-like enhancement of the irreducible susceptibility $\hat{\chi}_0(q)$.
\begin{subequations}
\label{eq:ch}
\begin{align}
\hat{\chi}^s(q) &=\hat{\chi}_0(q)+\hat{\chi}_0(q)\hat{U}^s\hat{\chi}^s(q), \\[3pt]
\hat{\chi}^c(q) &=\hat{\chi}_0(q)-\hat{\chi}_0(q)\hat{U}^c\hat{\chi}^c(q).
\end{align}
\end{subequations}
These susceptibilities contain all informations for not only spin and charge/orbital fluctuations but also higher-order multipolar fluctuations.
The specific matrix form of $\hat{\chi}_0(q)$ is denoted as~\cite{rf:order}
\begin{equation}
\hat{\chi}_0(q)=\!
\bordermatrix{
  & 11 & 21 & \cdots & 12 & \cdots  \cr
11& \chi^0_{11,11} & \chi^0_{11,21} &\cdots & \chi^0_{11,12} & \cdots \cr
21& \chi^0_{21,11} & \chi^0_{21,21} &\cdots & \chi^0_{21,12} & \cdots \cr
~\vdots &\vdots&\vdots&\ddots&\vdots& \cr
12 & \chi^0_{12,11} & \chi^0_{12,21} &\cdots & \chi^0_{12,12} & \cdots \cr
~\vdots &\vdots&\vdots&&\vdots& \ddots \cr
}
\end{equation}
With the use of the normal Green's functions, each component of the irreducible susceptibility is defined as follows:
\begin{subequations}
\begin{align}
&\chi^0_{\ell\ell',mm'}(q) =-\sum_k\mathcal{G}_{\ell m}(k+q)\mathcal{G}_{m'\ell'}(k) \\
&\label{eq:ch0}~~=\sum_{\bm r}\int_0^\beta\!d\tau \mathcal{G}_{\ell m}(\bm{r},\tau)\mathcal{G}_{\ell'm'}(\bm{r},\beta-\tau)e^{i\omega_n \tau-i\bm{k}\cdot\bm{r}}
\end{align}
\end{subequations}
The second line can be obtained with the use of the Fourier transformation and the relation of Eq.\eqref{eq:Gr}.
In these susceptibilities, a simple relation between the upper and lower triangular components holds.
This consideration also saves a memory in actual numerical calculations and speeds up the calculations.

In the FLEX approximation, first of all, we evaluate the eigenvalue $\epsilon_{n\bm k}$ and the unitary matrix $\tilde{u}^{\bm k}_{\ell m}$ for the unperturbed Hamiltonian, $\tilde{h}^{\bm k}_{\ell m}$ of Eq.\eqref{eq:hk}, and then, calculate the chemical potential $\mu$ by the condition that the electron density is a given $n$,
\begin{equation}
n=\sum_{n\bm{k}\sigma}f(\xi_{n\bm k})=2\sum_{n\bm k}f(\xi_{n\bm k}),
\end{equation}
where $\xi_{n\bm k}=\epsilon_{n\bm k}-\mu$, and $f(\epsilon)=1/(e^{\beta\epsilon}+1)$ is the Fermi-Dirac distribution function.
With the use of the obtained unitary matrix $\tilde{u}^{\bm k}_{\ell m}$, the noninteracting Green's functions are given by 
\begin{subequations}
\begin{align}
& \tilde{\mathcal{G}}^0_{\ell m}(\bm{k},i\omega_n)=\tilde{u}^{\bm k}_{\ell n}\tilde{u}^{\bm k}_{mn}\frac{1}{i\omega_n-\xi_{n\bm k}}, \\
& \tilde{\mathcal{G}}^0_{\ell m}(\bm{k},\tau)=-\tilde{u}^{\bm k}_{\ell n}\tilde{u}^{\bm k}_{mn}\big( 1-f(\xi_{n\bm k})\big) e^{-\xi_{n\bm k}\tau}.
\end{align}
\end{subequations}
We next transform $\tilde{\mathcal{G}}^0_{\ell m}(\bm{k},\tau)$ into $\mathcal{G}^0_{\ell m}(\bm{r},\tau)$, and then, evaluate Eqs.\eqref{eq:ch} via Eq.\eqref{eq:ch0}.
From Eqs.\eqref{eq:Sig} and \eqref{eq:Vq}, we can obtain the normal self-energy.
With Eqs.\eqref{eq:Gkt} and \eqref{eq:Sigt}, we obtain new normal Green's functions, determining a new chemical potential as 
\begin{equation}
n=2\sum_{n\bm k}f(\xi_{n\bm k})+2\sum_{m\bm k}\Big(\mathcal{G}_{mm}(\bm{k},i\omega_n)-\mathcal{G}^0_{mm}(\bm{k},i\omega_n)\Big).
\end{equation}
We carry out this self-consistent procedure until relative errors in $\tilde{\varSigma}_{\ell m}(k)$ becomes less than $10^{-4}$.
In this paper, we define the magnetic transition with $\chi^s(\bm{q},0) \ge 100$, where $\chi^s(q)$ is the spin susceptibility defined by
\begin{equation}
\chi^s(q)=\sum_{\ell m} \chi^s_{\ell\ell,mm}(q).
\end{equation}
Concerning the superconducting transition, from Eqs.\eqref{eq:Fk} and \eqref{eq:Dk}, we obtain the Eliashberg equation,
\begin{equation}
\label{eq:elia}
\begin{array}{l}
\displaystyle
\varDelta_{\ell m}(k)=-\lambda\sum_{k'}V^s_{\ell\ell'',m''m}(k-k') \\
\displaystyle
\hspace{45pt}\times\mathcal{G}_{\ell''\ell'}(k')\mathcal{G}_{m''m'}(k')^*\varDelta_{\ell'm'}(k')
\end{array}
\end{equation}
with the eigenvalue $\lambda$.
The transition temperature $T_{\rm c}$ can be obtained as the temperature when the maximum eigenvalue $\lambda$ is unity.


\begin{thebibliography}{99}
\bibitem{rf:diff} In the four band model in Ref.~\cite{rf:Arita}, the $(\pi,\pi)$ fluctuation favors the $d_{x^2-y^2}$-wave pairing, not $s_\pm$-wave.
The difference comes from inclusion of the $d_{3z^2-r^2}$-orbital.
As shown in Fig.~\ref{fig:qfs}(b), the presence of $d_{3z^2-r^2}$-orbital component near the Fermi level suppresses the $d_{x^2-y^2}$-wave pairing due to the high density of states.
\bibitem{rf:another} Although the forth band also holds the large gap amplitude on the inner $\alpha$ surface, the magnitude is the same as that on the outer $\alpha$ surface in the third band.
\bibitem{rf:order} We here define these matrix forms in different order from those in Ref.~\cite{rf:Takimoto}, due to easier treatment of actual calculations.
\end{thebibliography}
\end{document}